\documentstyle[12pt]{article}

\newcommand{\plb}[2]{{\em Phys. Lett.}          {\bf #1B}, #2 }
\newcommand{\npb}[2]{{\em Nucl. Phys.}          {\bf B#1}, #2 }
\newcommand{\pr }[2]{{\em Phys. Rep.}           {\bf  #1}, #2 }
\newcommand{\prd}[2]{{\em Phys. Rev.}           {\bf D#1}, #2 }
\newcommand{\prl}[2]{{\em Phys. Rev. Lett.}     {\bf  #1}, #2 }
\newcommand{\zpc}[2]{{\em Z. Phys.}             {\bf C#1}, #2 }
\newcommand{\sci}[2]{{\em Science}              {\bf  #1}, #2 }
\newcommand{\app}[2]{{\em Acta Phys. Polon.}    {\bf B#1}, #2 }
\newcommand{\mpl}[2]{{\em Mod. Phys. Lett.}     {\bf A#1}, #2 }
\newcommand{\cpc}[2]{{\em Comput. Phys. Commun} {\bf  #1}, #2 }
\newcommand{\con}[2]{                           {\bf  #1}, #2 }
\newcommand{\etal}{{\em et al.}}
\newcommand{\ibid}{{\em ibid.}}
\newcommand{\col}{Collaboration}
\newcommand{\ol}{\overline}
\newcommand{\ba}{\begin{array}}
\newcommand{\ea}{\end{array}}
\newcommand{\acht}{\\[8pt]}
\def\fiv{${\bf 5}+{\bf\ol5}$}
\def\ten{${\bf 10}+{\bf\ol{10}}$}

\def\ms{\mbox{{\footnotesize{$\overline{\rm MS}$}}} }
\def\lg{{\rm lg}}
\newcommand{\lsim}{\buildrel < \over {_\sim}}
\newcommand{\gsim}{\buildrel > \over {_\sim}}

\begin{document}
\thispagestyle{empty}
\begin{titlepage}
\title {\vspace{-2.0cm}
\hfill {\normalsize SLAC--PUB--7634} \\ \vspace{-0.3cm}
\hfill {\normalsize UPR--788--T} \\ \vspace{0.6cm}
Bounds on Supersymmetry from \\ Electroweak Precision Analysis}

\vspace{4.cm}

\author{ \\
{\sc Jens Erler}\thanks{e-mail: erler@langacker.hep.upenn.edu} \\ 
{\small Department of Physics and Astronomy}\\
{\small University of Pennsylvania}\\
{\small Philadelphia, PA 19104}\\
\\
{\sc Damien M. Pierce}\thanks{e-mail: pierce@slac.stanford.edu} \\ 
{\small Stanford Linear Accelerator Center} \\
{\small Stanford University} \\
{\small Stanford, CA 94309}}

\date{}
\maketitle
\thispagestyle{empty}

\begin{abstract}
\noindent
The Standard Model global fit to precision data is excellent. The
Minimal Supersymmetric Standard Model can also fit the data well,
though not as well as the Standard Model. At best, supersymmetric
contributions either decouple or only slightly decrease the total
$\chi^2$, at the expense of decreasing the number of degrees of
freedom. In general, regions of parameter space with large
supersymmetric corrections from light superpartners are associated
with poor fits to the data.  We contrast results of a simple (oblique)
approximation with full one-loop results, and show that for the most
important observables the non-oblique corrections can be larger than
the oblique corrections, and must be taken into account. We elucidate
the regions of parameter space in both gravity- and gauge-mediated
models which are excluded. Significant regions of parameter space are
excluded, especially with positive supersymmetric mass parameter
$\mu$. We give a complete listing of the bounds on all the
superpartner and Higgs boson masses. For either sign of $\mu$, and for
all supersymmetric models considered, we set a lower limit on the mass
of the lightest CP--even Higgs scalar, $m_h \geq 78$ GeV. Also, the
first and second generation squark masses are constrained to be above
280 (325) GeV in the supergravity (gauge-mediated) model.
\end{abstract}
\end{titlepage}

\setcounter{page}{1} 

\section{Introduction}

For more than a decade, ideas involving supersymmetry (SUSY) have been
among the most popular extensions of the Standard Model (SM), having
the greatest potential to solve its shortcomings such as the gauge
hierarchy problem and the lack of a quantized version of
gravity. Indeed, all superstring theories necessarily contain quantum
gravity and supersymmetry. Moreover, very recently, using arguments
based on various dualities and supersymmetry, all superstring theories
as well as 11 dimensional supergravity seem to be connected
nonperturbatively~\cite{Witten95}, suggesting a unified theory in
which supersymmetry is one of the indispensable key elements.  The
idea of supersymmetric unification is further supported by the
observation of gauge coupling unification~\cite{Dimopoulos81} two
orders of magnitude below the reduced Planck scale, $M_P/\sqrt{8 \pi}
\sim 2 \times 10^{18}$ GeV, the natural supergravity scale.

Taken together this implies a strong motivation to investigate the
phenomenological consequences of low energy supersymmetry. In this
paper we present a systematic study of precision observables in the
minimal supersymmetric standard model (MSSM) in an attempt to find
favorable and/or excluded regions in SUSY parameter space. We do this
in two scenarios for how supersymmetry breaking is conveyed to the
observable sector.

In the ``minimal supergravity'' model~\cite{Nilles84} supersymmetry
breaking is transmitted from a hidden sector to the observable sector
via gravitational interactions.  Supersymmetry is spontaneously broken
in the hidden sector at a scale $\Lambda$~\cite{Nilles83}. In popular
models which invoke gaugino condensation~\cite{Nilles82}, $\Lambda
\sim (M_P^2 M_{SUSY}/8 \pi)^{1/3} \sim 10^{13}$ GeV. Since gravity is
flavor blind, it is assumed that the explicit soft breaking terms in
the observable sector are universal at the supergravity scale.

In simple models of gauge mediation~\cite{Dine81} there is a
supersymmetry breaking sector which gives rise to both an $F$-term,
$F_X$, and a vacuum expectation value, $X$, of a standard model
singlet field.  This field is coupled to the vector-like ``messenger
fields", $M$, through a superpotential interaction of the form
$\lambda X M \overline{M}$. The generated soft mass term for a given
superpartner is proportional to $\Lambda\equiv F_X/X \sim 4\pi
M_Z/\alpha \sim 10^{5}$~GeV, and grows with the square of its gauge
couplings. Therefore sleptons are much lighter than squarks and the
gluino is heavier than charginos and neutralinos.

In both cases the generated soft terms are sufficiently flavor
universal that all flavor changing neutral current (FCNC) effects are
adequately suppressed. However, in the minimal supergravity model it
must be implicitly assumed that universality is not spoiled by the
K\"ahler potential.

Recently, a number of articles~\cite{Garcia95} examined whether
supersymmetry has the potential to describe the data better than the
SM. The conclusions in the affirmative were mainly driven by $R_b$
(see section~\ref{overview}), which at times was more than $3 \sigma$
higher than SM expectations. Loops involving light charginos and top
squarks could account for a large part of the discrepancy for low
$\tan\beta$, provided the effect was not canceled by charged-Higgs
loops.  For large $\tan\beta$, large shifts could be obtained from
loops containing bottom quarks and light neutral Higgs bosons, and to
a lesser extent from neutralino/bottom-squark loops. However, more
recent analyses find the measured value of $R_b$ closer to the SM
prediction (the discrepancy is $1.3 \sigma$), and at the same time
direct limits on superpartner masses have increased. Now, even if one
is able to find a region of parameter space where the $R_b$
discrepancy is alleviated, the decrease in the overall $\chi^2$ is not
significant. Also, in those regions of parameter space one typically
finds that the discrepancy in $A^{FB} (b)$ is made worse. In the
specific high-scale models of supersymmetry breaking we consider in
this paper, the largest possible shift in $R_b$ is less than $1
\sigma$. It follows that $R_b$ now plays a much smaller role, and
supersymmetric models can no longer yield significantly smaller values
of $\chi^2$ than the SM.

Therefore, in this work we take a different point of
view\footnote{First results of this kind of analysis were presented in
Ref.~\cite{Pierce97a}.} and focus on elucidating the excluded regions
of supersymmetry parameter space.  We present a complete one-loop
analysis of supersymmetry, combined with a state of the art SM
calculation. Input data are as of August 1997.

In Section~\ref{overview} we give a short overview of the inputs and
observables we use. For reference we perform a global fit to the SM,
and comment on some of the deviating observables.  Section~\ref{susy}
describes the parameter spaces of the supersymmetric models in more
detail, and reviews recent direct limits on superpartner masses. We
compare the oblique approximation~\cite{Peskin90} with the full
calculation in some detail. We present our results in
Section~\ref{results} and our conclusions in Section~\ref{con}.

\section{Overview}
\label{overview}

Within the SM we perform a global fit to a total of 31 observables
listed below. We include full one-loop radiative
corrections~\cite{Erler97}; full QCD corrections up to ${\cal O}
(\alpha_s^3)$; higher order QCD corrections when enhanced by beta
function effects; mixed electroweak/QCD corrections of ${\cal O}
(\alpha\alpha_s)$ with the exception of non-leading special vertex
corrections to $Z \rightarrow b\bar{b}$ decays\footnote{These have
been calculated very recently and shown to be quite small compared to
the analogous corrections to the lighter quark
vertices~\cite{Harlander97}.}; and ${\cal O} (\alpha^2)$ corrections
when enhanced, e.g. by large top mass effects or large logarithms.

The Fermi constant, $G_\mu$, and the fine structure constant,
$\alpha$, are taken as fixed inputs. The five fit parameters are the
strong coupling constant, $\alpha_s$; the contribution of the five
light quarks to the photon vacuum polarization function, $\Delta
\alpha_{\rm had}^{(5)}$; and the masses of the $Z$-boson, $M_Z$, the
Higgs scalar, $M_H$, and the top quark, $m_t$.  Alternatively, one may
fix $M_Z$ as well, since its relative error is now comparable to that
of $G_\mu$. We have chosen to leave it free, because the $Z$-mass
measurement is correlated with other observables. In practice, the
difference between the two treatments is numerically insignificant.
We have organized the measurements into seven groups:

\begin{itemize}


\item {\bf 9 lineshape observables}

The mutually correlated LEP observables~\cite{Abbaneo97} are
determined from a common fit to the $Z$ lineshape and the leptonic
forward-backward asymmetries. They include the $Z$-boson pole mass
$M_Z$, the total $Z$-width $\Gamma_Z$, the hadronic peak cross
section,
$$
\sigma_{\rm had}^0 = {12 \pi \over M_Z^2}{\Gamma_{ee} \Gamma_{\rm
had} \over \Gamma_Z^2}\ ,
$$
and, for each lepton flavor, $\ell = e$, $\mu$, $\tau$, the ratios
$R_{\ell} = \Gamma_{\rm had}/\Gamma_{\ell\ell}$, and the pole
asymmetries, $A^{FB} (\ell)$. $\Gamma_x$ denotes the partial $Z$ width
into $x$. Defining
$$
A_f = {1 - 4 Q_f \sin^2 \theta_{\rm eff}^f \over 
1 - 4 Q_f \sin^2 \theta_{\rm eff}^f + 8 Q_f^2 \sin^4 \theta_{\rm
eff}^f}\ ,
$$
where $\sin^2 \theta_{\rm eff}^f$ is the effective weak mixing angle
for fermion $f$ at the $Z$ scale, we have
$$
A^{FB} (f) = {3\over 4} A_e A_f\ .
$$


\item {\bf 3 further LEP asymmetries}

These are the $\tau$ polarization, ${\cal P} (\tau) = A_{\tau}$, its
forward-backward asymmetry, ${\cal P}^{FB} (\tau) = A_e$, and the
hadronic charge asymmetry, $\langle Q^{FB}\rangle$, which is quoted as
a measurement of $\sin^2 \theta_{\rm eff}^e$~\cite{Abbaneo97}.


\item {\bf 6 heavy flavor observables}

The mutually correlated heavy flavor observables from LEP and
SLC~\cite{Abbaneo97} are the ratios $R_q = \Gamma_{qq}/\Gamma_{\rm
had}$; the forward-backward pole asymmetries, $A^{FB} (q)$ (LEP); and
the combined left-right forward-backward asymmetries,
${A^{FB}_{LR}}(q) = A_q$ (SLC), each for $q = b, c$.


\item {\bf 3 further SLD asymmetries}

These are the very precise measurement of the left-right asymmetry for
hadronic final states~\cite{Abe97A}, $A_{LR}({\rm had}) = A_e$; the
analogous $A_{LR} (\mu,\tau) = A_e$, which is obtained from a common
fit including polarized Bhabha scattering~\cite{Abe97B}; the
left-right forward-backward asymmetries~\cite{Abe97B}, ${A^{FB}_{LR}}
(\ell) = A_\ell$ for $\ell = \mu$ or $\tau$; and the hadronic charge
flow asymmetry~\cite{Abe97C}, $A_Q = A_e$, which is basically the
ratio of weighted forward-backward and left-right forward-backward
asymmetries\footnote{Since $A_{LR} ({\rm had})$ is proportional to the
electron beam polarization, ${\cal P}_e$, while $A_Q$ is inversely
proportional, the geometric mean of the two determinations yields
$A_e$ independently of ${\cal P}_e$. The result~\cite{Abe97C} is $A_e
= 0.1574 \pm 0.0208$, or equivalently, $\sin^2 \theta_{\rm eff}^e =
0.2302 \pm 0.0027$.}. We combine the 3 values of $A_e$ from SLD into
one number,
$$ A_e = 0.1548 \pm 0.0033\ .$$


\item {\bf 2 pole masses}

We combine the $W$-mass measurements from CDF~\cite{Abe95},
D\O\ \cite{Abachi96} and UA2~\cite{Alitti92}, $M_W (pp) = 80.409 \pm
0.090$ GeV, with the one from LEP 2~\cite{Abbaneo97}, $M_W ({\rm LEP})
= 80.481 \pm 0.140$ GeV, obtaining
$$M_W ({\rm world}) = 80.430 \pm 0.076 \; {\rm GeV}.$$ Combining the
results of all top decay channels at CDF~\cite{Abe97D} and D\O\
\cite{Abachi97}, we find
$$m_t = 175 \pm 5 \; {\rm GeV}.$$ 
We fix the scale invariant \ms masses for the heavy quarks, 
$\ol{m}_b (\ol{m}_b) = 4.33 \; {\rm GeV}$ and
$\ol{m}_c (\ol{m}_c) = 1.30 \; {\rm GeV}$.


\item {\bf 6 low energy observables}

The weak charge from atomic parity violation (APV) in Tl has been
measured by groups in Oxford~\cite{Edwards95} and
Seattle~\cite{Vetter95}, $Q_W(^{205}{\rm Tl}) = - 114.77 \pm 1.23 \pm
3.44$, and in Cs by the Boulder group~\cite{Wood97}, $Q_W(^{133}{\rm
Cs}) = - 72.11 \pm 0.27 \pm 0.89$, where the first errors are
experimental and the second theoretical.  The recent result of the
deep inelastic scattering (DIS) experiment of CCFR~\cite{McFarland97}
is combined with the measurements of CDHS~\cite{Abramowicz86} and
CHARM~\cite{Allaby86}, yielding $\kappa = 0.5805 \pm 0.0039$, where
$\kappa$ is a linear combination of effective 4-Fermi operator
coefficients.  Similarly, $\nu_\mu e$ scattering experiments,
dominated by the recent CHARM II results~\cite{Vilain94}, yield
determinations of leptonic 4-Fermi operator coefficients, $g_V^{\nu e}
= -0.041\pm0.015$ and $g_A^{\nu e} = -0.507\pm0.014$.  The CLEO
measurement of $B\rightarrow X_s\gamma$~\cite{Alam95} yields the 90\%
confidence interval,
$$1\times10^{-4} < B(B\rightarrow X_s\gamma) < 4.2\times 10^{-4}.$$
When a variable is defined in a finite interval (here between 0 and
1), and a boundary of the interval is within a few standard deviations
from the central value, a variable transformation should be performed
such that the new variable is defined on the whole real axis.
Arguments analogous to the ones supporting a flat prior distribution
in $\log M_H$ rather than in $M_H$ (see next section) lead to the
logistic transformation\footnote{Equivalently, one may continue to use
the old variable and include a nontrivial weight factor (the Jacobian)
$1/[x(1-x)]$.}, $\lg(x)=\ln(x/(1-x))$~\cite{Gelman95}.  This results
in a more Gaussian shaped distribution, and in our case, in a more
(less) conservative assessment of the upper (lower) error bar.  The
$b\rightarrow s\gamma$ amplitude can receive very large contributions
from the leading-order supersymmetric loops, and the next-to-leading
order corrections (which have not been calculated) could be
important. The uncertainty in the large and positive MSSM
contributions is taken into account by the more conservative
treatment of the upper error bar. Additionally, because of these
uncertainties, we have chosen the more conservative of the two error
estimates in Ref.~\cite{Alam95}, which is obtained by adding the
statistical and systematic errors linearly.  Compared to this
experimental error, the theoretical error due to QCD uncertainties in
the SM prediction~\cite{Adel94} can be neglected.


\item {\bf 2 gauge couplings}

We use the constraint $\Delta \alpha_{\rm had}^{(5)} = 0.02817 \pm
0.00062$~\cite{Alemany97}.  We also include the external constraint
$\alpha_s = 0.118 \pm 0.003$, which we obtained by combining non-$Z$
lineshape determinations of $\alpha_s$~\cite{Burrows97}.  In this case
we have doubled all theoretical errors to account for ignored
correlations, and because these are by far the most difficult to
estimate and very non-Gaussian in nature.


\end{itemize}

The best fit values of the SM input parameters are shown in
Table~\ref{fit}.  The $\chi^2/{\rm d.o.f.}$ of the fit is 26.6/26,
corresponding to an almost perfect fit with a goodness (probability of
larger $\chi^2$) of 43\%.
\begin{table}[htb]
\begin{center}
\begin{tabular}{|l|c|} \hline
$m_t$                           & $172 \pm 5    $ GeV \\
$M_H$                           & $66^{+74}_{-38}      $ GeV \\
$M_Z$                           & $91.1867  \pm 0.0020 $ GeV \\
$\alpha_s (M_Z)$                & $0.1195  \pm 0.0022 $ \\
$\Delta\alpha_{\rm had}^{(5)}$  & $0.02817  \pm 0.00064$ \\
\hline
\end{tabular}
\end{center}
\caption{Results for the SM fit parameters.}
\label{fit}
\end{table}
Our results differ from the ones quoted by the LEP Electroweak Working
Group~\cite{Abbaneo97}. The differences can be traced to a different
treatment of radiative corrections and to a slightly different and
more recent data set.  Our results are in close correspondence with
the very recent update of the Particle Data Group~\cite{Erler97}; one
major difference is the external $\alpha_s$ constraint. Also, here
$M_H$ is allowed as a free parameter, in which case the SM best fit
value of $M_H$ is below the lower limit from direct searches. In the
MSSM the bound is relaxed to $m_h \gsim 60$ GeV~\cite{Sopczak97},
consistent with the central value in Table~\ref{fit}.

We list the measured values of all the observables in Table~\ref{obs},
together with their global best fit values. The pull, defined as the
difference between the measurement and the fit result divided by the
experimental error, is also shown. The agreement is excellent.  The
two largest discrepancies are only at the $2 \sigma$ level.
\begin{table}[htb]
\begin{center}
\begin{tabular}{|l|c|c|r|} \hline
 & measurement & SM & pull \\ 
\hline \hline
$M_Z$ [GeV] &          $91.1867 \pm 0.0020$ & 91.1867 &    0.0 \\
$\Gamma_Z$ [GeV] &     $ 2.4948 \pm 0.0025$ &  2.4959 & $-0.4$ \\
$\sigma_{\rm had}$[nb]&$41.486  \pm 0.053 $ & 41.478  &    0.2 \\
$R_e$ &                $ 20.757 \pm 0.056 $ & 20.744  &    0.2 \\
$R_\mu$ &              $ 20.783 \pm 0.037 $ & 20.744  &    1.1 \\
$R_\tau$ &             $ 20.823 \pm 0.050 $ & 20.789  &    0.7 \\
$A^{FB} (e)$ &         $ 0.0160 \pm 0.0024$ &  0.0163 & $-0.1$ \\
$A^{FB} (\mu)$ &       $ 0.0163 \pm 0.0014$ &  0.0163 &    0.0 \\
$A^{FB} (\tau)$ &      $ 0.0192 \pm 0.0018$ &  0.0163 &    1.6 \\
\hline                      
${\cal P} (\tau)$ &    $ 0.1411 \pm 0.0064$ &  0.1476 & $-1.0$ \\
${\cal P}^{FB} (\tau)$&$ 0.1399 \pm 0.0073$ &  0.1476 & $-1.1$ \\
$\sin^2 \theta_{\rm eff}^e(Q^{FB})$ &
                       $ 0.2322 \pm 0.0010$ &  0.2315 &    0.8 \\
\hline
$R_b$ &                $ 0.2170 \pm 0.0009$ &  0.2158 &    1.3 \\
$R_c$ &                $ 0.1734 \pm 0.0048$ &  0.1722 &    0.2 \\
$A^{FB} (b)$ &         $ 0.0984 \pm 0.0024$ &  0.1035 & $-2.1$ \\
$A^{FB} (c)$ &         $ 0.0741 \pm 0.0048$ &  0.0739 &   0.0  \\
$A_{LR}^{FB} (b)$ &    $ 0.900  \pm 0.050 $ &  0.935  & $-0.7$ \\
$A_{LR}^{FB} (c)$ &    $ 0.650  \pm 0.058 $ &  0.668  & $-0.3$ \\
\hline
$A_e$ &                $ 0.1548 \pm 0.0033$ &  0.1476 &    2.2 \\
$A_{LR}^{FB} (\mu)$ &  $ 0.102  \pm 0.034 $ &  0.148  & $-1.3$ \\
$A_{LR}^{FB} (\tau)$ & $ 0.195  \pm 0.034 $ &  0.148  &    1.4 \\
\hline 
$M_W$ [GeV] &          $80.430  \pm 0.076 $ & 80.386  &    0.6 \\
$m_t$ [GeV] &          $ 175    \pm 5     $ & 172     &    0.6 \\
\hline
$Q_W ({\rm Cs})$ &     $-72.11  \pm 0.93  $ &$-73.11$ &    1.1 \\
$Q_W ({\rm Tl})$ &     $-114.8  \pm 3.7   $ &$-116.7$ &    0.5 \\
$\kappa ({\rm DIS})$ & $ 0.581  \pm 0.0039$ & 0.583   & $-0.7$ \\
$g_V^{\nu e}$ &        $-0.041  \pm 0.015 $ &$-0.0396$& $-0.1$ \\
$g_A^{\nu e}$ &        $-0.507  \pm 0.014 $ &$-0.5064$& $ 0.0$ \\
$\lg(B(B\rightarrow X_s\gamma))$ &
		       $ -8.49  \pm 0.45  $ & $-7.99$ & $-1.1$ \\  
\hline
$\Delta\alpha_{\rm had}^{(5)}$ &
                       $ 0.02817\pm 0.00062$& 0.02817 &    0.0 \\
$\alpha_s (M_Z)$ &     $ 0.118  \pm 0.003 $ & 0.1195  & $-0.5$ \\
\hline \hline
\end{tabular}
\end{center}
\caption{Results of a global fit to the Standard Model. For each
observable, we list the experimental result, the best fit result
within the SM with the Higgs mass allowed as a free parameter, and the
pull.}
\label{obs}
\end{table}

Based on all data from 1992--1996, the left-right asymmetry,
$A_{LR}({\rm had}) = 0.1550\pm 0.0034$, is now closer to the Standard
Model expectation of $0.147\pm 0.002$ than previously~\cite{Abe97A}.
However, because of the smaller error, $A_{LR}$ is still significantly
above the Standard Model prediction. There is also an experimental
difference of $\sim 1.9 \sigma$ between the SLD value of $A_\ell\;
({\rm SLD}) = 0.1547 \pm 0.0032$ from all $A_{LR}$ and $A_{LR}^{FB}
(\ell)$ data on one hand, and the LEP value $A_{\ell}\; ({\rm LEP}) =
0.1461\pm 0.0033$ obtained from $A^{FB} (\ell)$, $A_e({\cal P}_\tau)$,
and $A_\tau({\cal P}_\tau)$ on the other hand, in both cases assuming
lepton-family universality.

Relaxing universality, one can extract the various $A_f$ in a model
independent way. First, we combine $A_e$ from $A^{FB}(e)$ and ${\cal
P}^{FB} (\tau)$ with the value from the SLC to obtain $A_e = 0.1518
\pm 0.0029$. This value for $A_e$ can now be used to extract the other
$A_f$ from forward-backward asymmetries. Looking at Table~\ref{obs}
one sees that all 3 observables which are sensitive to the
$\tau$-vertex, namely $A^{FB} (\tau)$, ${\cal P}(\tau)$, and
$A_{LR}^{FB}(\tau)$, deviate by $1 \sigma$ or more. However, when
combined, one finds $A_\tau = 0.1448 \pm 0.0059$, in excellent
agreement with the SM value of $0.147 \pm 0.002$. At present, it seems
more likely that we are facing an experimental discrepancy. Using
$A^{FB} (f)$ from LEP and $A^{FB}_{LR}(f)$ from the SLC for $f = \mu$,
$b$, and $c$, we find that $A_\mu$ and $A_c$ are consistent with each
other and the SM.  On the other hand, while the $A_b$ determinations
are also consistent with each other, their combined value, $A_b =
0.872 \pm 0.024$, deviates $2.6 \sigma$ from the SM value of 0.935.  A
similar analysis assuming lepton universality yields~\cite{Abbaneo97}
$A_b = 0.877 \pm 0.023$ which is $2.5 \sigma$ low.  One can also
obtain $A_e = 0.148 \pm 0.002$ from a global fit excluding the quark
sector\footnote{Here we have fixed $M_H = M_Z$ because otherwise the
Higgs mass is driven to very low (excluded) values.}, and derive $A_b
= 0.888 \pm 0.022$, still a $2.1 \sigma$ deviation. We can therefore
conclude that the data show an anomaly in $A_b$, independently of
exactly how it is extracted. However, this deviation of about 6\%
cannot be due to supersymmetric loops since a 30\% correction to
$\hat\kappa_b$ (defined by $\sin^2 \theta_{\rm eff}^b = \hat\kappa_b
\sin^2 \hat\theta_{\overline{\rm MS}}$) would be necessary to account
for the central value of $A_b$.  Only a new contribution at tree level
which does not contradict $R_b$ (including the off-peak $R_b$
measurements by DELPHI~\cite{Abreu95}), can conceivably account for
such a low $A_b$~\cite{Erler95}.  In particular, a $\tau$-sneutrino
with mass close to $M_Z$ and R-parity violating couplings is a natural
candidate~\cite{Erler97A}.  Of course, supersymmetry can also affect
$A^{FB}(b)$ through the weak mixing angle in the $A_e$ factor. Such a
correction can not, however, help to resolve the LEP/SLC $A_\ell$
discrepancy.

This discussion and the remarks about $R_b$ in the introduction show
that we cannot expect that the inclusion of supersymmetric loops will
significantly improve the goodness of the fit to electroweak data. On
the contrary, there are regions in parameter space where the fit is
quite poor, and those regions must be ruled out. Here we do not test
the supersymmetry hypothesis. Rather, the primary goal of our analysis
must be to identify the excluded regions of supersymmetry parameter
space.

\section{Supersymmetry}
\label{susy}

In gravity-mediated models supersymmetry is broken in a hidden sector
and the breaking is communicated to the observable sector through
supergravity effects. This transmission is suppressed by inverse
powers of the Planck mass. The assumption of a flat K\"ahler metric
for the fields in the observable sector (an assumption which can be
called into question) yields a universal scalar mass, $M_0$, a
universal gaugino mass, $M_{1/2}$, and a universal trilinear scalar
coupling, $A_0$, at the Planck scale~\cite{Girardello82}.  In the
minimal supergravity model we impose this universality at the scale
where the (U(1) and SU(2)) gauge couplings unify, $M_{\rm
GUT}\sim2\times10^{16}$ GeV. The running of the soft parameters from
$M_P$ to $M_{\rm GUT}$ should in general be taken into
account~\cite{Polonsky94}, but it is model dependent, so it is ignored
in the minimal supergravity model.

{}From the inputs $M_Z$, $G_\mu$, $\alpha$, $\Delta\alpha_{\rm
had}^{(5)}$, and $\alpha_s$, and the fermion masses $m_t$, $m_b$, and
$m_\tau$, we determine the full one-loop values of the gauge and
Yukawa couplings at the $Z$-scale.  We then solve the set of coupled
two-loop renormalization group equations (RGE's) for the couplings and
soft parameters by iteration.  This set of differential equations has
two-sided boundary conditions: the hard parameters (gauge and Yukawa
couplings) are fixed at the $Z$-scale and the soft parameters are
fixed at the unification scale.  In each iteration we impose
electroweak symmetry breaking at the squark scale. This occurs
naturally, as the large top-quark Yukawa coupling drives the Higgs
mass-squared $m_{H_2}^2$ negative. The electroweak symmetry breaking
conditions allow us to solve for the Higgsino mass-squared, $\mu^2$,
and the CP--odd Higgs boson mass-squared, $m_A^2$. These are
determined as functions of $\tan\beta$, the ratio of vacuum
expectation values, to full one-loop order~\cite{Pierce97}.  To
summarize, the minimal supergravity model parameter space is specified
by
\begin{equation}
\label{su ps}
M_0,\ M_{1/2},\ A_0,\ \tan\beta,\ {\rm and\ sgn}(\mu)\ .
\end{equation}

Just as in minimal supergravity, simple models of gauge-mediated
supersymmetry breaking also introduce a hidden sector in which
supersymmetry is dynamically broken, so that for one (or more) hidden
sector fields $X$, $F_X \neq 0$.  Here, however, the messenger fields
which couple to $X$ are assumed to be charged under the SM gauge
group. As a result, supersymmetry breaking is transmitted through
messenger loop effects, and observable fields receive soft masses
roughly proportional to the square of their gauge couplings. It
follows that scalars with identical quantum numbers are mass
degenerate, and FCNC effects are sufficiently suppressed as long as
the messenger scale $M = \lambda X \lsim 10^{15}$
GeV~\cite{Dimopoulos95}.  The key parameter here is $\Lambda = F_X/X$,
as this parameter sets the scale of the entire superpartner spectrum.
The messenger scale $M$ determines the amount of RGE evolution, and
therefore enters only logarithmically. As before, radiative
electroweak symmetry breaking occurs rather generically over the
parameter space. It is generally difficult to arrange for the right
magnitude of $\mu$ and the soft supersymmetry breaking parameter $B$,
but solutions to this problem exist~\cite{Dine96}. We assume that the
mechanism which is responsible for generating $B$ and $\mu$ does not
at the same time give extra contributions to the soft squark and
slepton masses.  The superpartner spectrum depends on the
representation(s) of the messenger sector. We consider the cases where
the messenger sector is comprised of a \fiv \ pair of fields, or a
\ten\ pair\footnote{The messenger fields are assumed to come in
complete $SU(5)$ representations to preserve the successful gauge
coupling unification. An actual $SU(5)$ symmetry, however, is not
assumed.}. For the purpose of our analysis a \ten\ messenger sector is
equivalent to a model with three \fiv\ pairs.  In the minimal
gauge-mediated model the supersymmetry parameter space includes
\begin{equation}
\label{gm ps} M,\ \Lambda,\ \tan\beta,\ {\rm and\ sgn}(\mu),
\end{equation}
with $\Lambda < M$ to ensure the absence of gauge symmetry breaking in
the messenger sector.

In either supersymmetry model, we proceed by randomly picking a point
in its parameter space, Eq.~(\ref{su ps}) or Eq.~(\ref{gm ps}).  For
an initial set of values for the {\em standard parameters\/}, $M_Z$,
$m_t$, $\alpha_s$ and $\Delta\alpha_{\rm had}^{(5)}$, we then solve
the model, i.e.\ we self-consistently evaluate the soft parameters at
the weak scale, and compute the physical superpartner and Higgs-boson
masses, couplings and mixings.  This enables us to compute the full
supersymmetric one-loop
corrections~\cite{Pierce97,Grifols84,Boulware91} to each
observable\footnote{For the low-energy observables from atomic parity
violation, deep inelastic scattering, and neutrino scattering, we work
in the oblique approximation.  For these observables the non-oblique
supersymmetric corrections are expected to be negligible relative to
the experimental errors.}, and to perform a $\chi^2$ minimization with
respect to the standard parameters.  We iterate the solution of the
supersymmetric model for the best fit values of the standard
parameters until this process has converged.  On the last iteration we
require that the full one-loop electroweak symmetry breaking
conditions are satisfied, and impose that the Yukawa couplings remain
perturbative ($\lambda^2/4\pi < 1$) up to $M_{\rm GUT}$.

Finally, we apply the direct lower limits on superpartner and Higgs
boson masses. Searches at LEP constrain the mass of the CP--odd Higgs
boson, $A$, to be larger than 62.5 GeV, and that of the lightest
CP--even Higgs scalar, $h$, to be heavier than 62.5 GeV for small
$m_A$, or $70.7$ GeV for large $m_A$~\cite{AlephHEP97}. There are mass
limits from LEP 2 on charginos ($m_{\tilde\chi^+} >
91$~GeV)~\cite{Janot97}, selectrons ($m_{\tilde e} > 76$~GeV), smuons
($m_{\tilde\mu} > 59$~GeV), staus ($m_{\tilde\tau} >
53$~GeV)~\cite{Barate97}, and stops ($m_{\tilde t} >
65$~GeV)~\cite{Cerutti97} provided these superpartners are not nearly
degenerate with the lightest neutralino, $\tilde\chi_1^0$.  The
chargino limit is slightly degraded if the electron sneutrino is
light.  The top squark limit is for the case when it decouples from
the $Z$, and increases to 72 GeV for no left-right mixing. For a
relatively light $\tilde\chi_1^0$, D\O\ excludes top squarks up to 95
GeV~\cite{Abachi96A}.  In the gauge-mediated models the
next-to-lightest supersymmetric particle (NLSP) can be a slepton. We
assume in these models that the NLSP decays outside the detector, in
which case the search limits for stable charged particles apply, which
are $m_{\tilde\ell} > 67$ (69)~GeV for right-handed (left-handed)
sleptons~\cite{Barate97A}. The for us relevant CDF
searches~\cite{Abe97E} suggest for first and second generation
squarks, $m_{\tilde q} > 210$~GeV, and for gluinos, $m_{\tilde g} >
173$, but these limits do not apply uniformly over parameter
space. Therefore we impose relaxed limits of 200 and 160 GeV,
respectively. However, it turns out that in the models we consider the
direct limits on squark and gluino masses are for the most part
irrelevant. Once the other mass limits are taken into account, the
smallest possible squark (gluino) masses vary from 210 (230) to 370
(330) GeV, depending on the model.

At each point in our scan over parameter space we apply the
constraints from electroweak symmetry breaking, Yukawa perturbativity,
and the direct searches. If any of these checks fail, the point is
disregarded. We refer to a point in the supersymmetry parameter space
which passes these tests as a {\em quaint point}. The set of points
which pass comprise the {\em quaint parameter space}.

We now discuss the interpretation of the resulting $\chi^2$ value for
a given quaint point.  It is important to notice that the overall
$\chi^2$ value is relevant only for hypothesis testing. As noted
before, this is not our goal. We derive our confidence that the study
of low-energy supersymmetry is worthwhile from other (more indirect)
arguments. Rather, we will address the question of preferred or
disfavored quaint regions. Before we continue to defend our procedure,
we would like to compare our situation with the SM Higgs mass.

In standard analyses, the quoted limits on $M_H$ are independent of
the $\chi^2$ value of the SM best fit. A standard (Bayesian) procedure
is to determine the {\em posterior probability distribution
function\/} (posterior p.d.f.)  as the product of the {\em prior
p.d.f.}\ (prior) and the likelihood function (here $\sim
\exp(-\chi^2/2))$. The choice of a prior is always made, but is
oftentimes obscured. In the case of $M_H$, a {\em non-informative\/}
prior is desired, or perhaps an informative prior which accounts for
the direct lower limit, and is non-informative otherwise. The choice
of a suitable prior is not unique\footnote{The sensitivity of the
posterior p.d.f.\ to the (non-informative) prior diminishes rapidly
with the inclusion of more data.}.  A non-informative prior is by no
means necessarily flat (constant); in general a flat prior ceases to
be constant after a variable transformation introduces a non-trivial
Jacobian. Clearly, such a variable transformation cannot change the
results of a statistical analysis unless the Jacobian is dropped. In
the case of $M_H$, one often chooses a prior which is flat in $\log
M_H$.  There are various ways to justify this
choice~\cite{Gelman95}. One rationale is that a flat distribution is
most natural for a variable defined over the real numbers. This is the
case for $\log M_H$ but not $M_H^2$.  Another complication arises
since flat priors in both $\log M_H$ and $M_H^2$ are improper, i.e.\
their integrals are divergent and cannot be normalized.  This is not
necessarily problematic as long as the likelihood function is
sufficiently convergent so that the posterior p.d.f.\ is
normalizable. For $M_H$ this is the case, due to the non-decoupling
property of the SM Higgs boson corrections.

In the case of supersymmetry, however, the posterior p.d.f.\ remains
improper, because superpartners and the heavy Higgs doublet decouple,
and the likelihood function is asymptotically finite. To cope with
this problem one might
\begin{itemize}
\item 
cut superparticle masses or soft parameters off somewhere in the TeV
region, arguing that higher scales are unnatural and incompatible with
the primary motivations for low-energy supersymmetry such as the
hierarchy problem. In other words one would introduce an informative
prior. However, conclusions would directly depend on the precise
choice of the cut-off scale, i.e.\ the prior;
\item 
choose a more sophisticated informative prior which utilizes
fine-tuning criteria in order to produce a proper posterior
p.d.f. While aesthetically appealing in particular from a Bayesian
viewpoint\cite{Bayes1763}, the implementation would be rather involved
and computationally expensive. Also the establishment of an
appropriate fine-tuning criteria is not unique~\cite{Barbieri88};
\item
exclude all quaint points which have a $\Delta \chi^2 = \chi^2 -
\chi^2_{\rm min} > 3.84$, where $\chi^2_{\rm min}$ is the minimum
value of $\chi^2$ in the supersymmetry model under consideration. In a
univariate Gaussian situation this procedure corresponds to the 95\%
CL. It lacks the conceptual rigor of Bayesian statistics, but has the
advantage of being independent of the specification of prior
information. We will use this criterion in our analysis.
\end{itemize}

Before showing the results of the full one-loop analysis, we discuss
the results obtained in the oblique approximation.  For some
observables the dominant supersymmetric corrections are of oblique
type, i.e.\ they arise from vector-boson self-energy diagrams and can
be described by the $S$, $T$ and $U$ parameters\footnote{We use the
definitions from reference~\cite{Marciano90}, but redefine them to
include contributions from new physics
only~\cite{Erler95a}.}~\cite{Peskin90}.  These parameters can only be
extracted from the data when $M_H$ is fixed to a reference value. The
following approximate formulas reproduce $S$, $T$ and $U$ within $\pm
0.01$ for 60 GeV $< M_H <$ 1000 GeV:
\begin{equation}
\label{stu}
\ba{l} 
S = - 0.12 \pm 0.14 + 0.04 ({100\; {\rm GeV}\over M_H} - 1) 
                    - 0.13 \log_{10} {M_H\over 100\; {\rm GeV}},\acht
T = - 0.10 \pm 0.14 + 0.08 ({100\; {\rm GeV}\over M_H} - 1) 
                    + 0.31 \log_{10} {M_H\over 100\; {\rm GeV}},\acht
U = + 0.20 \pm 0.24 - 0.02 ({100\; {\rm GeV}\over M_H} - 1).  
\ea
\end{equation}
We see that all 3 oblique parameters are consistent with zero within
$1 \sigma$.  Now consider the supersymmetric contributions to $S$, $T$
and $U$. Supersymmetry contributes to the $S$ parameter with either
sign, but that changes the overall $\chi^2$ only by relatively small
amounts typically of ${\cal O}(1)$.  Contributions to $T$ and $U$ are
positive definite, but $U$ is increased by less than 0.1, which is
much less than the present uncertainty. On the other hand,
contributions to $T$ can be significant, and since positive,
Eq.~(\ref{stu}) shows that strong constraints can be expected to be
associated with $T$.

\begin{figure}[htb]
\vbox{\kern12.5cm\includegraphics{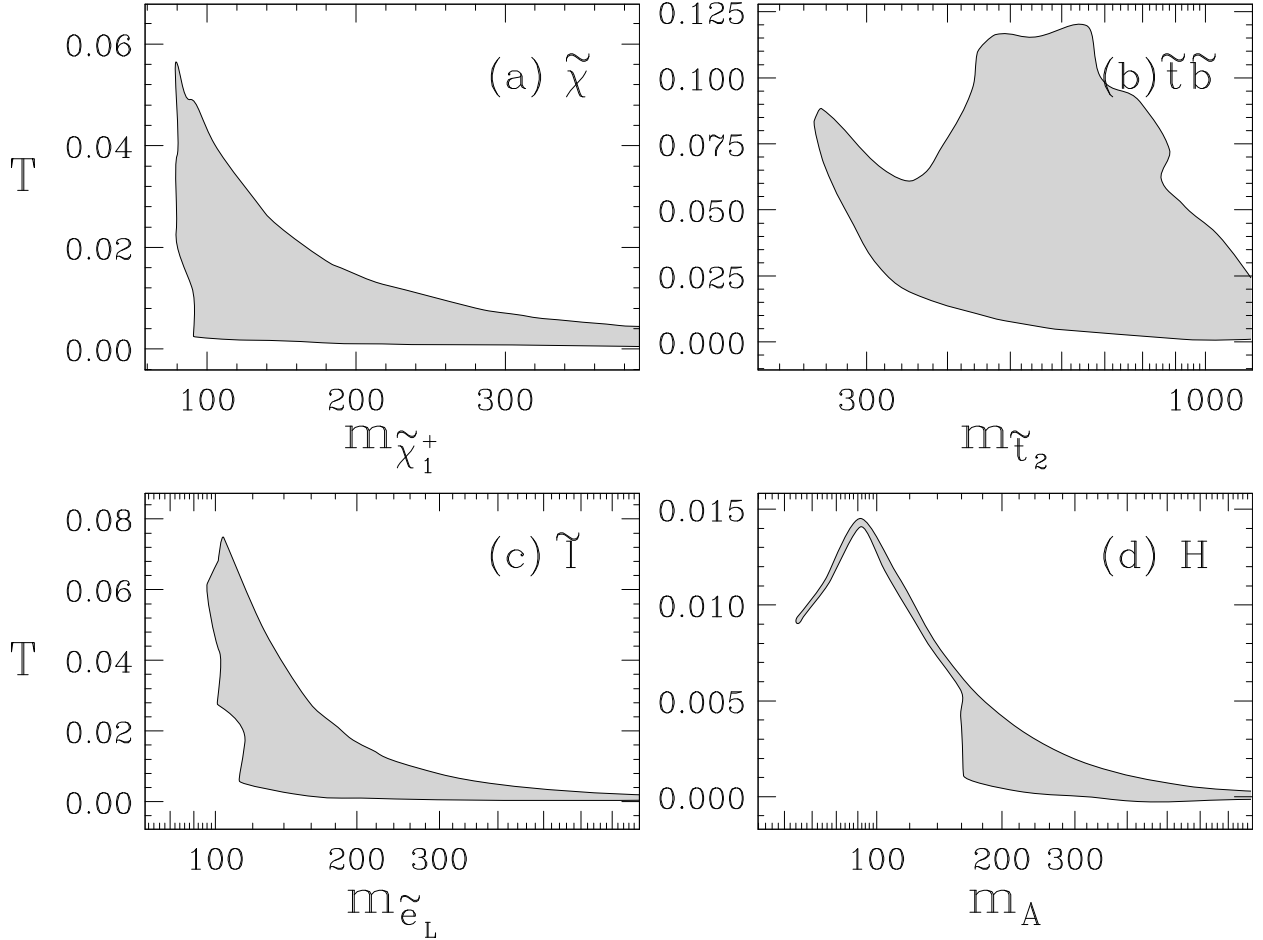}}
\caption{Supersymmetric contributions to the $T$ parameter in the
minimal supergravity model. Shown are
(a) the chargino/neutralino contributions vs.\ the light chargino mass;
(b) the stop/sbottom contributions vs.\ the heavy stop mass; 
(c) the slepton contributions vs.\ the left-handed selectron mass; and 
(d) the Higgs sector contributions vs.\ the CP--odd Higgs boson mass.
Masses are in GeV.}
\label{Tpar}
\end{figure}

The general oblique fit in Eqs.~(\ref{stu}) is not very suitable for
supersymmetry.  It is more appropriate to impose the constraints $T >
0$, $U > 0$, and 60 GeV $< M_H <$ 150 GeV, and to deviate from the
oblique approximation in allowing an extra parameter, $\gamma_b$,
defined through~\cite{Altarelli93A} $\Gamma(Z\to b \bar{b}) = \Gamma^0
(Z\to b \bar{b}) (1+\gamma_b)$, in order to account for the
possibility of supersymmetric $Zb\bar{b}$ vertex corrections.  The
result is\footnote{This fit does not include the constraint from
$B(B\rightarrow X_s\gamma)$.}

\begin{equation}
\label{stub}
\ba{l} 
S = - 0.09_{-0.10}^{+0.18}, \\
T \leq + 0.06, \\
U = + 0.17_{-0.17}^{+0.23}, \\
\gamma_b = 0.006 \pm 0.004.
\ea
\end{equation}

If we redo this fit with $\gamma_b = 0$, the strong constraint, $T
\leq 0.06$ is relaxed to $T \lsim 0.08$. Eq.~(\ref{stub}) shows that
the $T$ parameter is the most important of the three oblique
parameters in determining the goodness of the fit.  We show the
various supersymmetric contributions to the $T$ parameter in
Fig.~\ref{Tpar} in the supergravity model. The shaded regions are
determined by randomly choosing ${\cal O} (10^5)$ points in parameter
space (\ref{su ps}) to find 50,000 quaint points. We see that the
maximal contributions from each of the three superpartner sectors,
gauginos, third generation squarks, and sleptons are significant
($\delta T_{\rm max} \sim$ 0.06, 0.12, and 0.08, respectively).  The
Higgs sector (and the first two generations of squarks) contributes at
a lower level ($\delta T_{\rm max}\sim$ 0.015). This pattern holds for
the $S$ and $U$ parameters as well.

\begin{figure}[htb]
\vbox{\kern6.cm\includegraphics{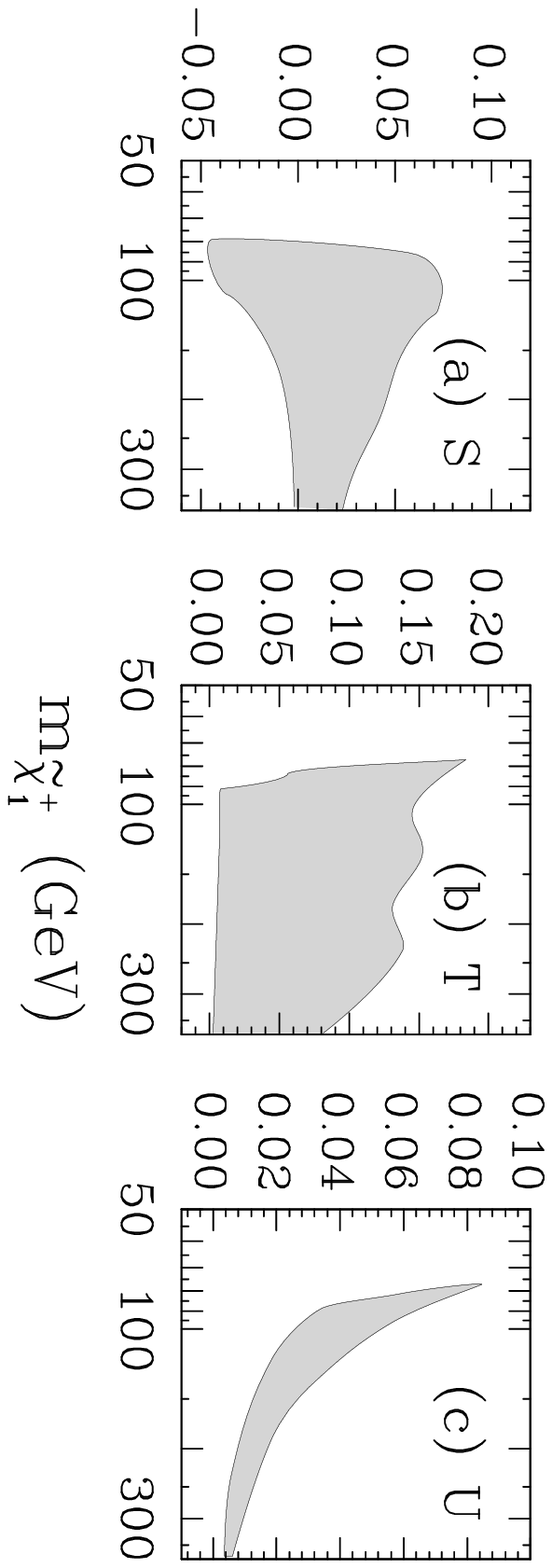}}
\caption{Supersymmetric contributions to the $S$, $T$, and $U$
parameters vs. the light chargino mass, in the supergravity model.}
\label{STUpar}
\end{figure}

\begin{figure}[htb]
\vbox{\kern6.cm\includegraphics{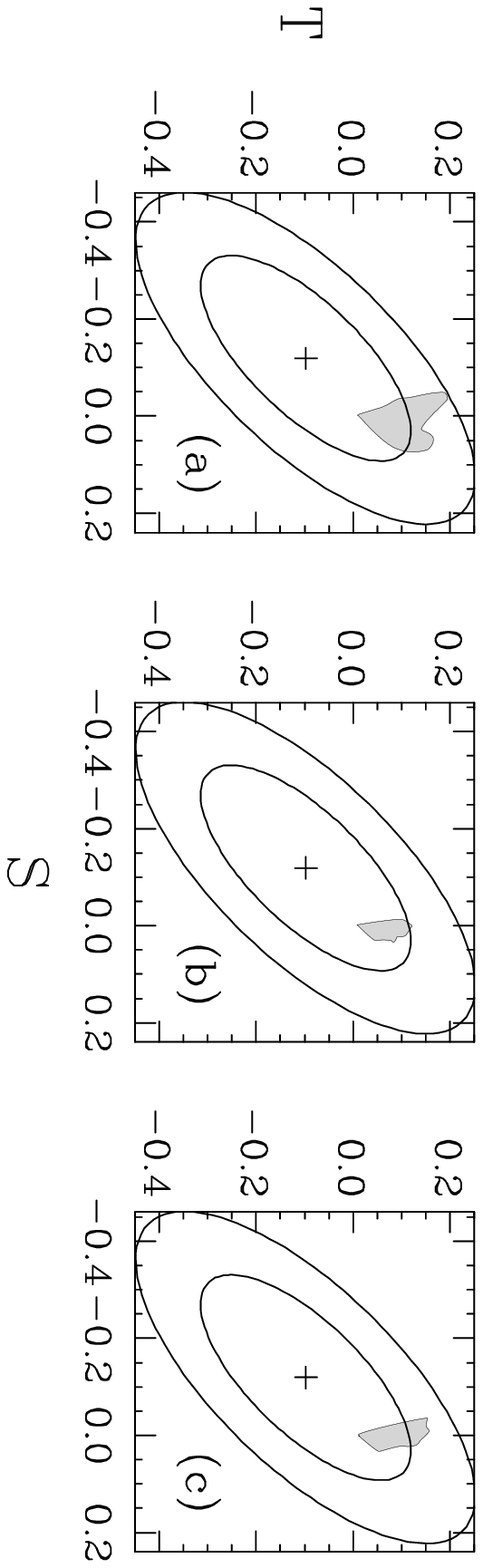}}
\caption{Supersymmetric contributions to the $S$ and $T$ parameters.
Shown are the results for (a) the minimal supergravity model; (b) the
\fiv\ gauge-mediated model; and (c) the \ten\ gauge-mediated model.
The best fit value ({\tt +}) and 68 and 95\% CL contours of $S$ and
$T$ are also shown, for $M_H=100$ GeV. Note that by definition $S = T
= 0$ in the SM.}
\label{STpar}
\end{figure}

We show the total contribution from all superpartners and extra Higgs
bosons to $S$, $T$ and $U$ in Fig.~\ref{STUpar}. Here we see that
these contributions are approximately in the range ($-0.05$ to
$+0.08$), (0 to 0.2), and (0 to 0.09), respectively, for $S$, $T$ and
$U$, and the decoupling of supersymmetry from these parameters is
evident.  These contributions are significantly limited by the recent,
more stringent collider bounds on new particle masses discussed above.
In the gauge-mediated models, the maximum contribution to the oblique
parameters is even further reduced.  We illustrate this in
Fig.~\ref{STpar}, which shows the allowed regions in the $S$--$T$
plane for the three models. We also show the best fit values of $S$
and $T$, and the 68\% and 95\% contours, determined with $M_H$ fixed
at 100~GeV.  It is important to realize that the difference in
$\chi^2$ between the SM minimum and the minimum allowing freely
varying oblique parameters is only one unit.  Hence, even with the
best possible addition of oblique corrections, the $\chi^2$ will not
improve significantly with respect to the SM. Moreover,
Fig.~\ref{STpar} nicely illustrates two points. First, the corrections
from supersymmetry are quite limited in magnitude, and second, they
tend to worsen the fit compared to the SM.

\begin{figure}[htb]
\vbox{\kern13cm\includegraphics{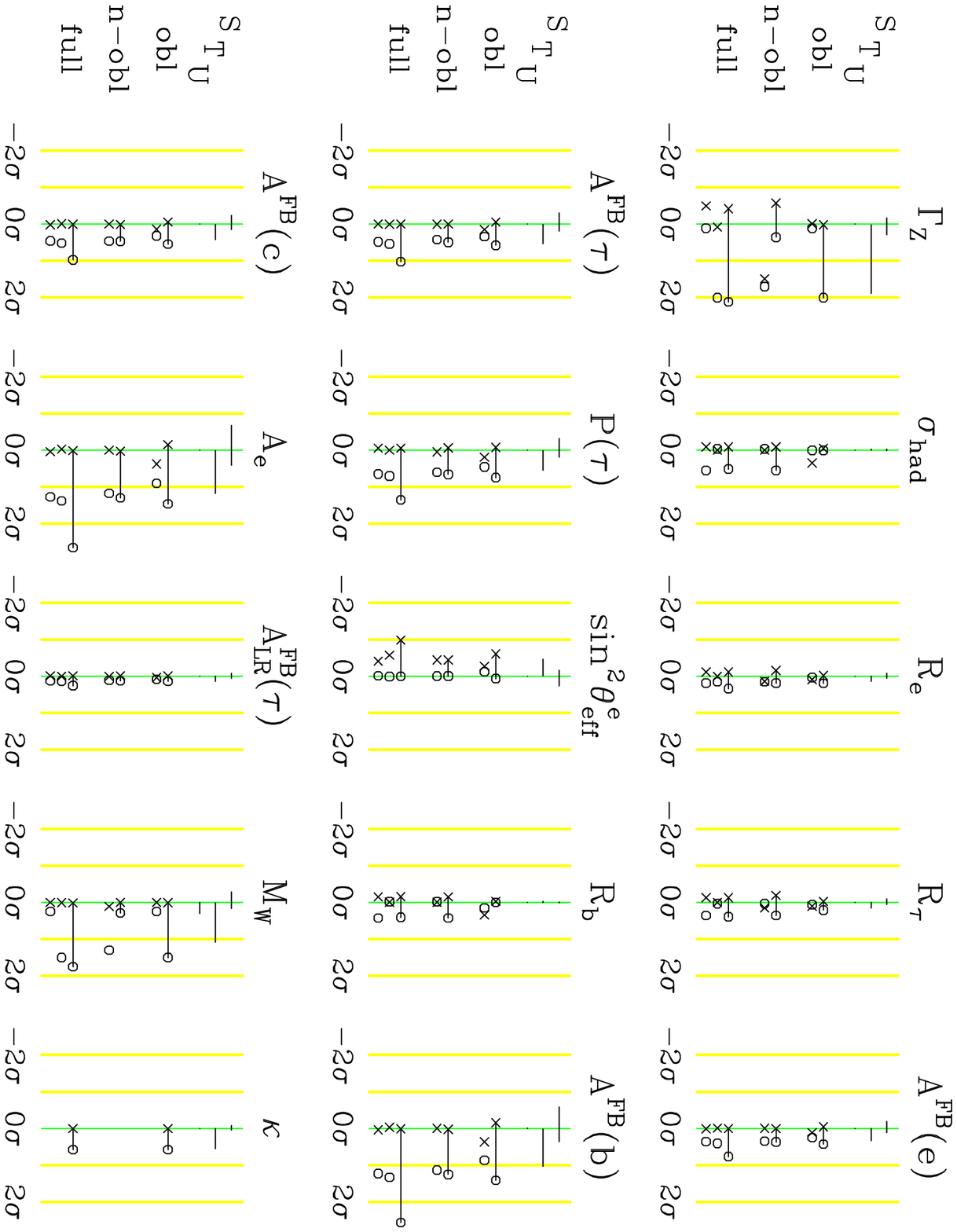}}
\caption{Supersymmetric contributions to various precision observables
in the supergravity model. Shown is the effect on each observable (in
units of standard deviation) when each oblique parameter, $S$, $T$, or
$U$, is varied within its possible range in the supergravity
model. Similarly, we show the full range of the combined effect of all
three oblique parameters (labeled {\tt obl}), so that a circle (cross)
corresponds to the point with the largest positive (negative) oblique
shift for that observable. Just below, we indicate the size of the
corresponding non-oblique corrections at these points. Likewise, the
full range of the non-oblique corrections are shown (labeled {\tt
n-obl}), and just below the corresponding oblique corrections at the
extrema of this range are indicated. Finally, we show the range of the
full corrections found in our scan (labeled {\tt full}), and at the
endpoints of this range we break the corrections down into their
oblique and non-oblique parts. These are indicated just below,
respectively. The non-oblique corrections to $\kappa$ are not shown.}
\label{sugraobs}
\end{figure}

\begin{figure}[htb]
\vbox{\kern13cm\includegraphics{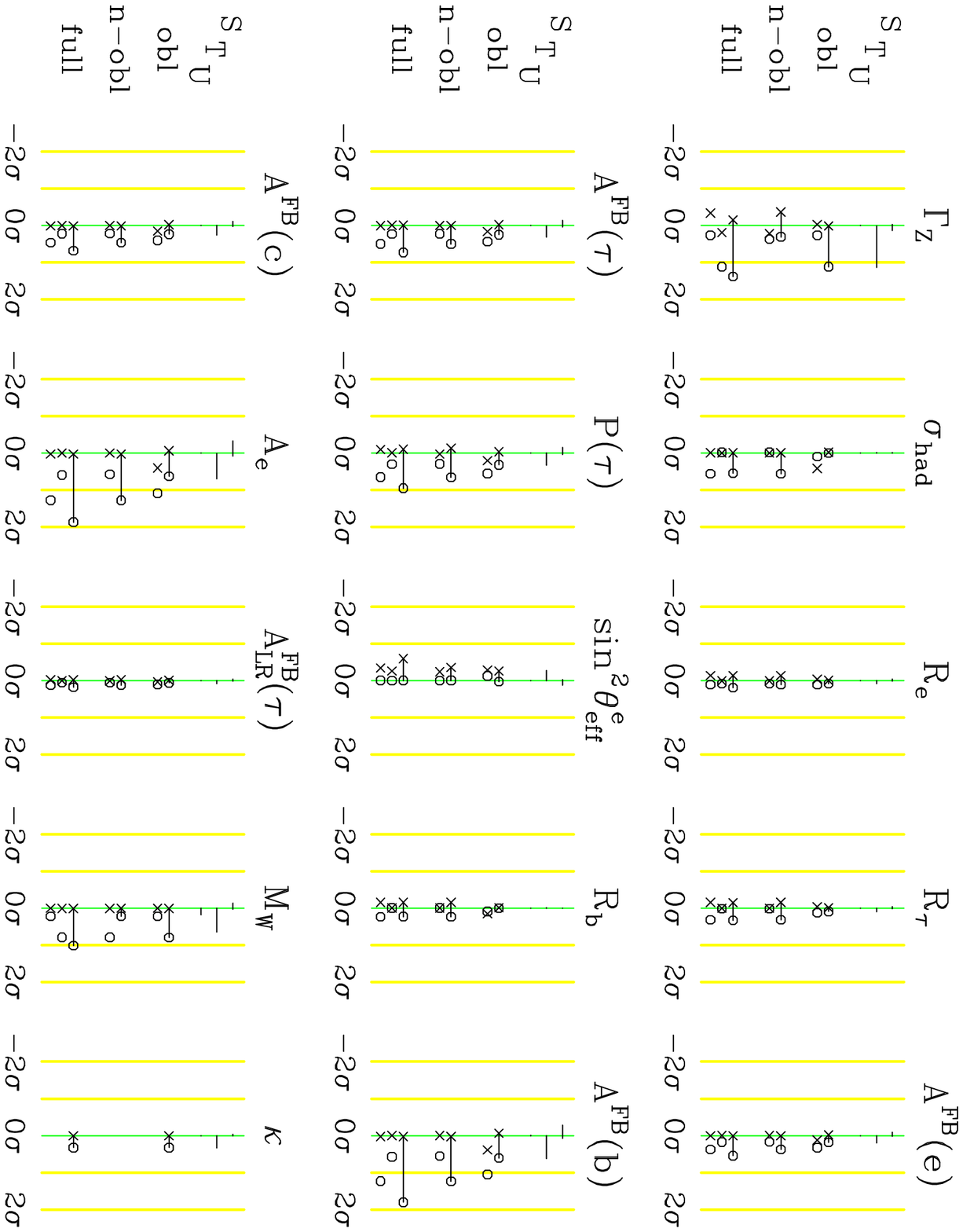}}
\caption[]{As Fig.~\ref{sugraobs}, for the \fiv\ gauge-mediated model.}
\label{gm5obs}
\end{figure}

\begin{figure}[htb]
\vbox{\kern13cm\includegraphics{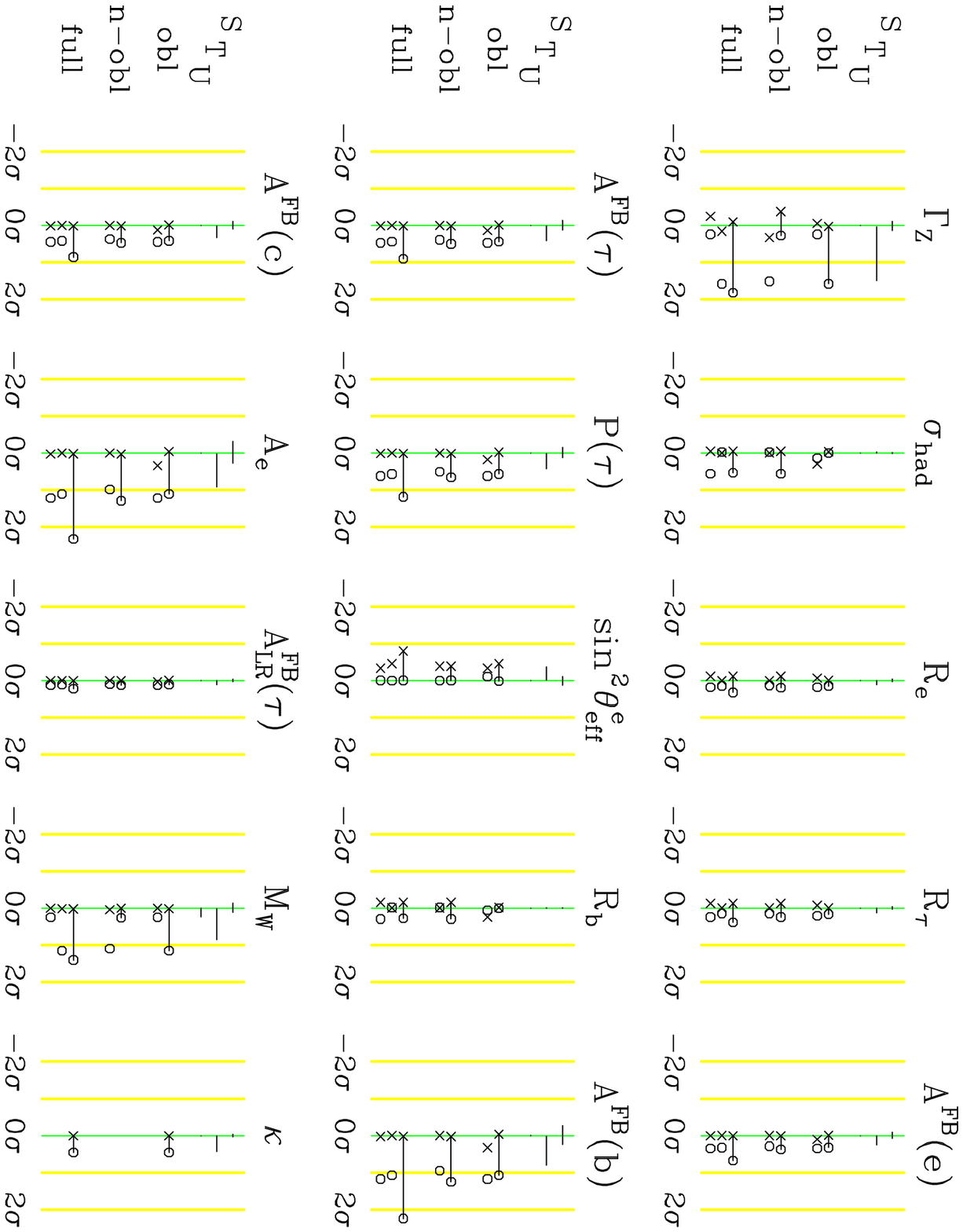}}
\caption[]{As Fig.~\ref{sugraobs}, for the \ten\ gauge-mediated model.}
\label{gm10obs}
\end{figure}

Judging from Fig.~3, it appears that no or very little of the
supersymmetric parameter space can be excluded. However, the oblique
approximation is just that, an approximation. As we will show, it
cannot be used to reliably indicate whether parameter space is
excluded. The oblique approximation neglects higher order terms in the
derivative expansion, as well as the non-universal corrections: the
vertex corrections, wave-function renormalization, and box diagrams.
We illustrate the relative magnitude of the oblique and non-oblique
corrections to many of the observables in Fig.~\ref{sugraobs}, in the
supergravity model. Here, we show the full range of the shifts of each
observable due to the shift in each oblique parameter\footnote{Only
$M_W$ receives non-negligible contributions from the $U$ parameter. In
particular, the DIS variable $\kappa$ is insensitive to $U$, although
DIS results are frequently quoted in terms of $\sin^2\theta_W = 1 -
M_W^2/M_Z^2$.  Clearly, in the presence of new physics such quotes can
be misleading.}, labeled {\tt S, T, U}, and their sum\footnote{The
shift from the maximal total oblique correction is not necessarily the
sum of the shifts due to the individual maximal oblique corrections,
because the individual corrections need not find their maximal values
at the same point in parameter space.}, labeled {\tt obl}. For
comparison, just below the range of the shift from the total oblique
correction we indicate the non-oblique corrections at the endpoints of
that range.  Therefore one can compare the oblique and non-oblique
corrections at the points of maximal positive and negative oblique
shift. We also show the range of the non-oblique corrections in our
scan, labeled {\tt n-obl}, and just below that, for comparison, the
corresponding oblique correction at the points of maximal positive and
negative non-oblique shift. Finally, we show the range of the full
one-loop corrections, labeled {\tt full}, and at the points of maximal
positive and negative full correction we indicate the relative sizes
of the oblique and non-oblique parts. All these shifts are shown in
units of standard deviation, i.e.\ we divide each shift by the
experimental error. The APV observables, the neutrino scattering
observables, as well as $A^{FB}_{LR} (\mu)$, $A^{FB}_{LR} (b)$,
$A^{FB}_{LR} (c)$, and $R_c$, all receive very small shifts, and are
not shown.  We show the same plots for the \fiv\ and \ten\ gauge
mediation models in Figs.~\ref{gm5obs} and \ref{gm10obs}. From these
figures we see that, in all three models, the non-oblique corrections
can be larger than the oblique corrections {\em for every observable
except $M_W$} (and $\Gamma_Z$ in the supergravity model). Hence it is
essential to include the full corrections in this type of analysis.

Figs.~\ref{sugraobs}--\ref{gm10obs} also show that the largest
corrections occur in the supergravity model, and the smallest in the
\fiv\ gauge-mediated model. In each model the gaugino masses obey the
usual GUT relation, i.e. $M_i\propto\alpha_i$. For a given gaugino
mass scale, the scalar spectra (and the Higgsino mass) in the three
models vary. The model with the lightest scalars (for a fixed gaugino
mass) is the supergravity model, in the region $M_0 \ll M_{1/2}$.  It
is in this region (actually $M_0 \sim M_{1/2}/2$) that the largest
corrections occur. The \fiv\ gauge-mediated model has the heaviest
scalar spectrum, and hence the smallest corrections
(Fig.~\ref{gm5obs}). The scalar masses can be made relatively lighter
in the gauge-mediated case if the effective number of \fiv\ pairs,
$n_5^{\rm eff}$, is increased. The \ten\ model we consider corresponds
to $n_5^{\rm eff}=3$, and hence the corrections are larger in the
\ten\ case (Fig.~\ref{gm10obs}). However, they are still not as large
as in the supergravity model. We expect the magnitude of the
corrections in the case\footnote{This corresponds to an SO(10) model
with a ${\bf 16} + \ol{\bf 16}$ pair of fields.} $n_5^{\rm eff}=4$ to
match quite closely with the supergravity results, as in this case the
two spectra can have significant overlap (see Figure~10, Table~1, and
the corresponding discussion in Ref.~\cite{Bagger97}, for an example).
Figs.~\ref{gm5obs} and~\ref{gm10obs} show that insensitive observables
in the supergravity model are likewise insensitive in the \fiv\ and
\ten\ gauge mediation models.

\begin{figure}[htb]
\vbox{\kern20cm\includegraphics{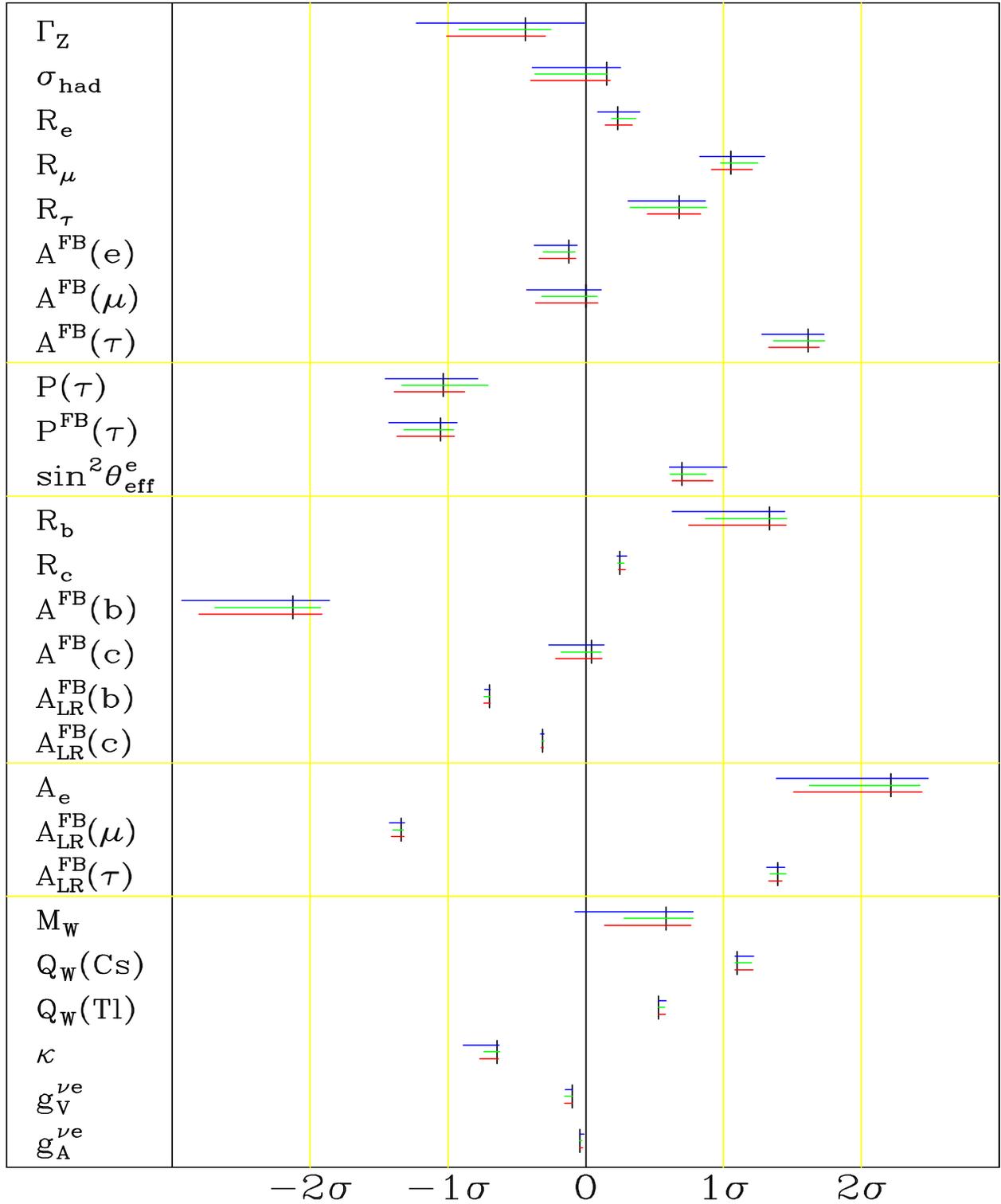}}
\caption{The range of best fit predictions of precision observables in
the supergravity model (upper horizontal lines), the \fiv\
gauge-mediated model (middle lines), the \ten\ gauge-mediated model
(lower lines), and in the Standard Model at its global best fit value
(vertical lines), in units of standard deviation.}
\label{allobs}
\end{figure}

Figure~\ref{allobs} shows, for each model, the range of the best fit
predictions of each observable found in our scan over quaint parameter
space, in units of standard deviation. We also indicate the best fit
value of each observable in the SM. Note that the range of each
observable in the supersymmetric models always includes the SM best
fit value, since the supersymmetric corrections decouple, and the
range of the supersymmetric light Higgs boson mass includes the best
fit SM Higgs mass. The best fit values of the observables which are
most sensitive to supersymmetric contributions, $\Gamma_Z$,
$A^{FB}(b)$, and the SLC measurement of $A_e$, can vary over a
range of $\sim 1 \sigma$. This range for the best fit predictions is
smaller than the actual contributions from supersymmetry shown in
Figs.~\ref{sugraobs}--\ref{gm10obs}, which can be larger than $2
\sigma$.  This is because the standard parameters adjust themselves to
minimize the overall $\chi^2$. The sensitivity of $\Gamma_Z$,
$A^{FB}(b)$, and $A_e$ are due to the very high precision with
which these observables are measured. Shifts in $M_W$ are somewhat
smaller, but still important. Taken together, these large shifts do
not improve but rather worsen the fit. They give rise to regions in
parameter space which can be excluded. We describe these excluded
regions in the next section. Not shown in Fig.~\ref{allobs} are
correlations among the supersymmetric corrections. These correlations
figure importantly in the fit.  For example, in the region of
parameter space where the supersymmetric prediction for $R_b$ is
closest to the measured value, the discrepancy in $A^{FB}(b)$ is much
worse.

\section{Results}
\label{results}

\begin{figure}[htb]
\vbox{\kern15.cm\includegraphics{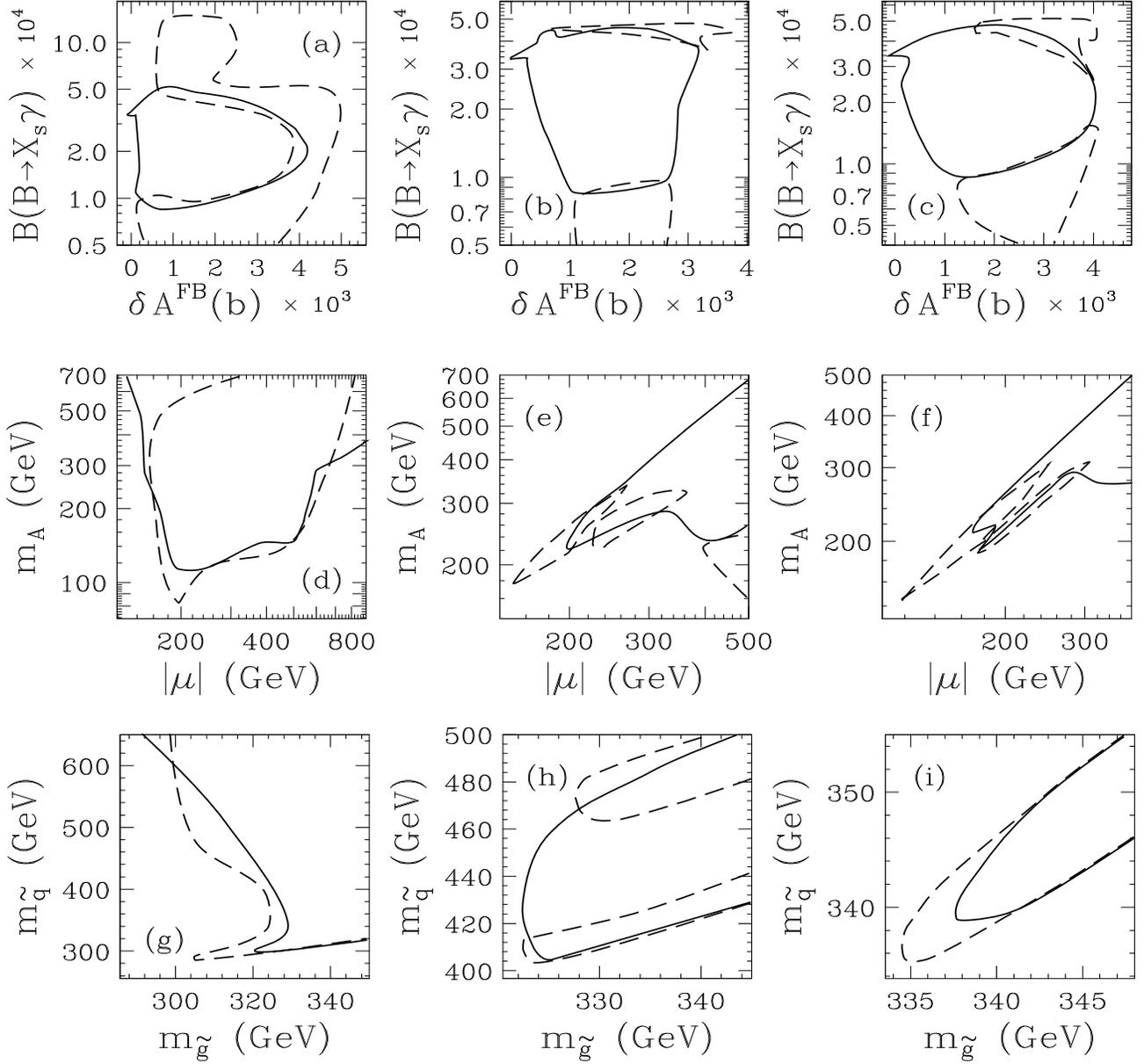}}
\caption{Allowed and excluded regions for various observables and
supersymmetry parameters in the minimal supergravity model, and the
\fiv and \ten\ gauge-mediated models (respectively, from left to
right), with $\mu < 0$. The regions which contain allowed points lie
inside the solid curves, while the regions which contain excluded
points are bounded by dashed curves.  Therefore, the regions inside
dahsed curves and outside solid curves are absolutely excluded,
independent of the values of other parameters.}
\label{neg1}
\end{figure}

\begin{figure}[htb]
\vbox{\kern14.cm\includegraphics{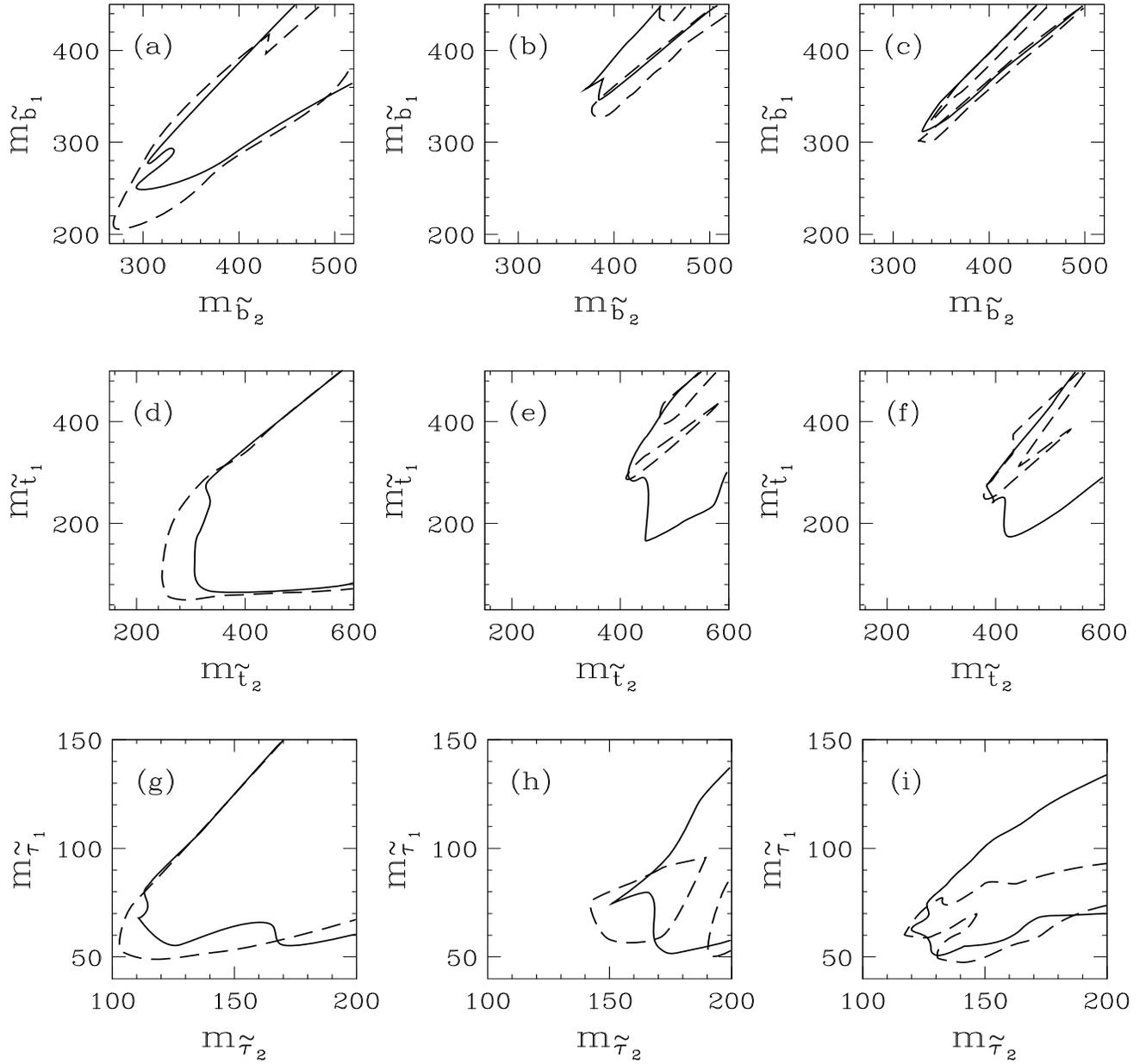}}
\caption[]{Allowed and excluded $\mu<0$ regions in third generation
squark and slepton mass parameter spaces, as described in
Fig.~\ref{neg1}. Masses are in GeV.}
\label{neg2}
\end{figure}

\begin{figure}[htb]
\vbox{\kern22.cm\includegraphics{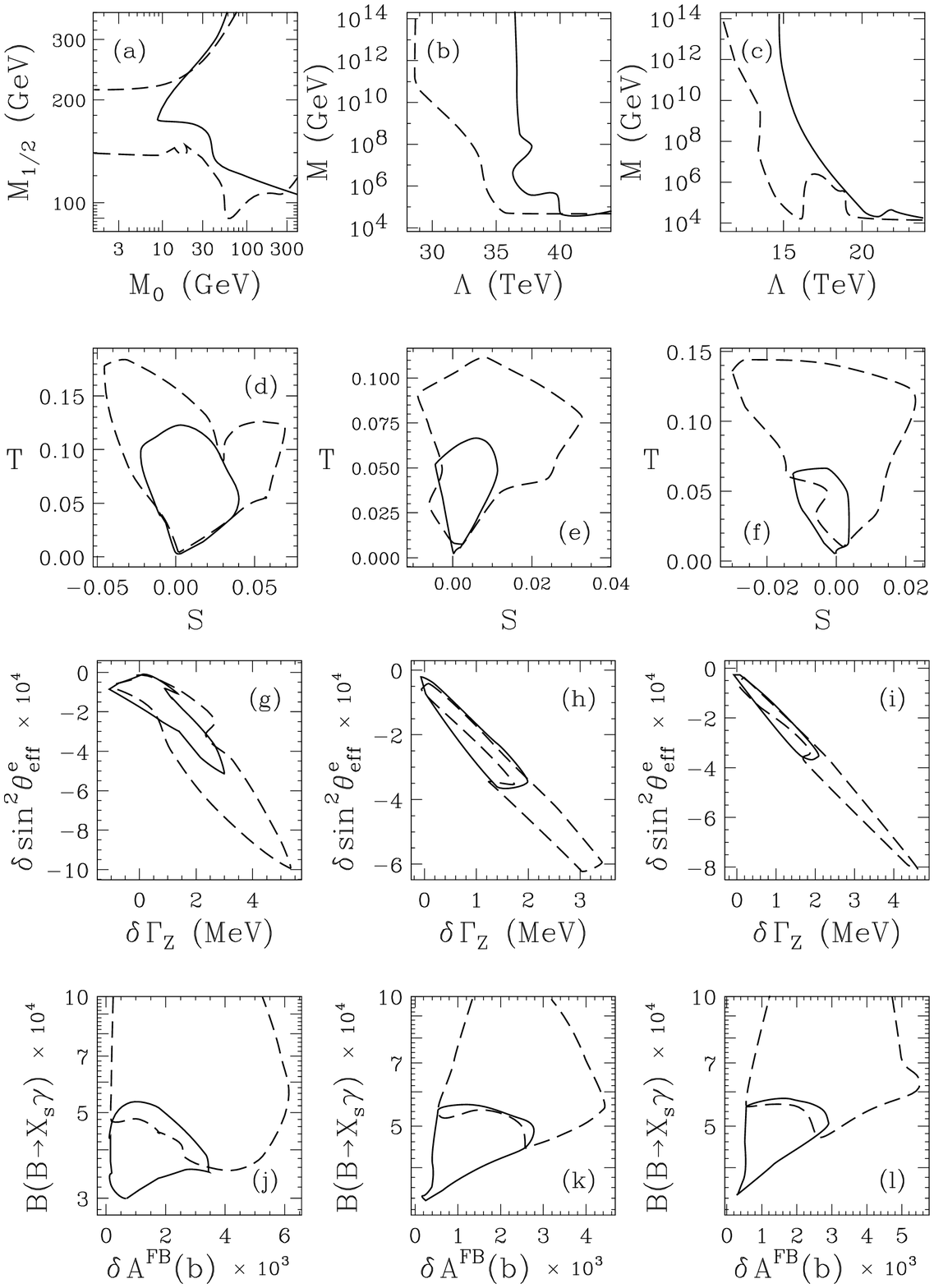}}
\caption[]{Allowed and excluded regions, as in Fig.~\ref{neg1}, with
$\mu>0$.}
\label{pos1}
\end{figure}

\begin{figure}[htb]
\vbox{\kern22.cm\includegraphics{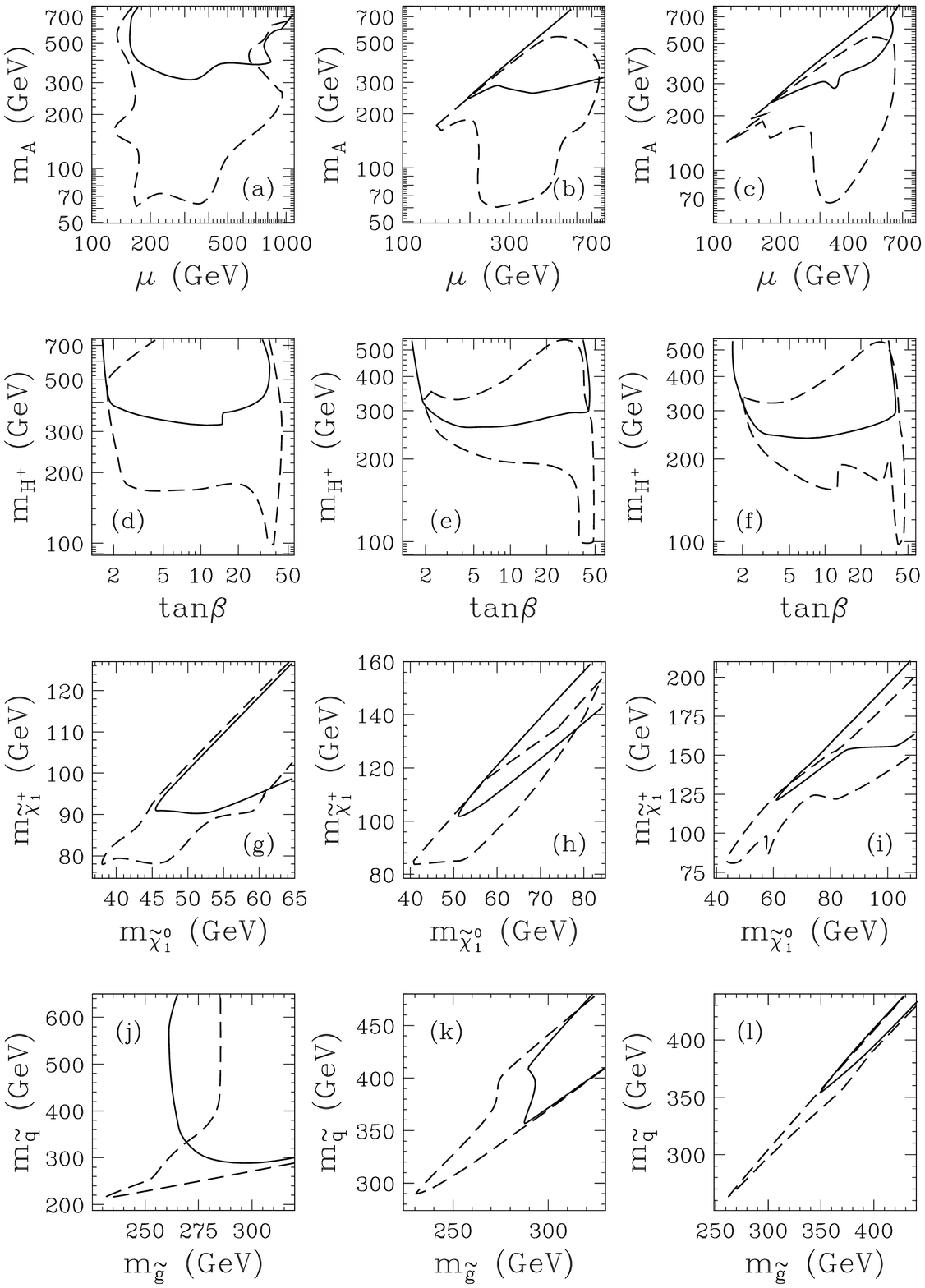}}
\caption[]{Allowed and excluded regions, as in Fig.~\ref{neg1}, with
$\mu>0$.}
\label{pos2}
\end{figure}

\begin{figure}[htb]
\vbox{\kern22.cm\includegraphics{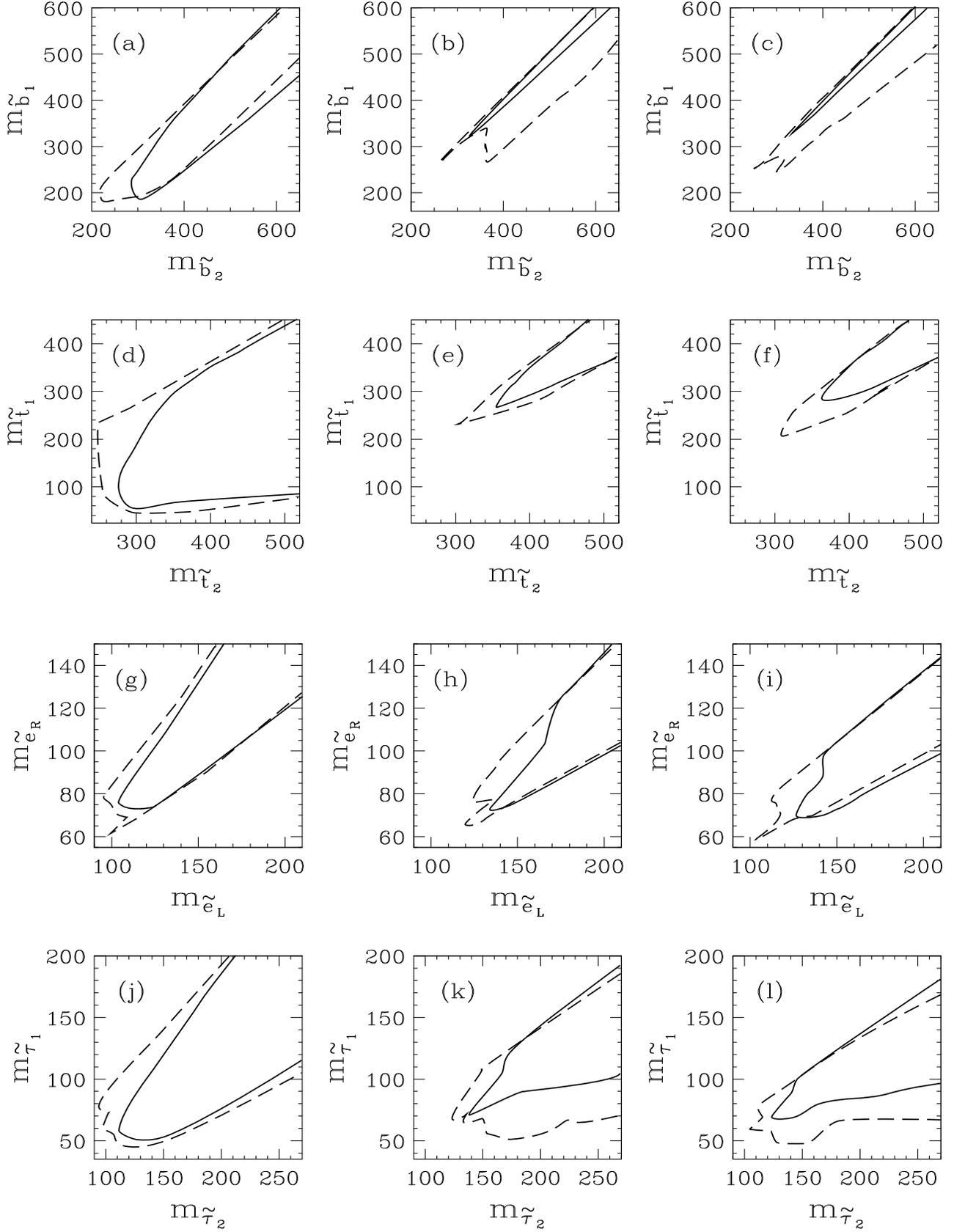}}
\caption[]{As Fig.~\ref{neg1}, in squark and slepton mass parameter
spaces, with $\mu>0$. Masses are in GeV.}
\label{pos3}
\end{figure}

The allowed and excluded quaint regions in a variety of
two-dimensional parameter subspaces are shown in Figs.~\ref{neg1} and
\ref{neg2} for $\mu < 0$, and in Figs.~\ref{pos1}, \ref{pos2}, and
\ref{pos3} for $\mu > 0$.  These plots are the main result of our
analysis.  Each row of these figures shows the results for the
supergravity model, the \fiv\ gauge-mediated model, and the \ten\
gauge-mediated model, respectively from left to right. We show with a
solid line the region of parameter space where it is possible to find
$\Delta\chi^2 < 3.84$.  With a dashed line we indicate the region
where it is possible to find $\Delta \chi^2 > 3.84$. Hence, those
regions enclosed by a dashed line but not by a solid line are
excluded, independently of the value of any other parameter. Of
course, in the regions bounded by a dashed line {\it and\/} a solid
line there are both allowed and excluded points. In some cases we plot
the shift $\delta$ from supersymmetry rather than the value of the
observable. We show more results in the $\mu>0$ case because of the
larger excluded region, as discussed below. In the following
discussion, we use the abbreviations SUGRA, GM$_5$, and GM$_{10}$ to
refer to the minimal supergravity, \fiv\, and \ten\ gauge-mediated
models. Also, GM refers to {\em both\/} the \fiv\ and \ten\ models,
and limits listed as three consecutive numbers refer to the SUGRA,
GM$_5$, and GM$_{10}$ models, respectively.

Figs.~\ref{neg1} (a--c) and \ref{pos1} (j--l) demonstrate the
importance of the $B(B\rightarrow X_s\gamma)$ observable for our
results\footnote{Since the next-to-leading order calculation of
$B(B\rightarrow X_s\gamma)$ is not available in the MSSM, we use an
improved leading order formula in the Standard Model calculation
\cite{Cho91}, and include the leading order supersymmetric corrections
\cite{Bertolini91}.}.  When included much larger values of
$\Delta\chi^2$ are possible. Large parts of parameter space are
excluded due to unacceptably large positive or negative contributions
to the $b \rightarrow s \gamma$ amplitude, especially with $\mu >
0$. The most important corrections are due to chargino/top-squark
loops. These contributions are proportional to $\mu\tan\beta$, at
large $\tan\beta$. Hence, for $\mu > 0$, they constructively add to
the SM and charged Higgs amplitudes, and very large total amplitudes
can result over a wide region of parameter space.  For large negative
$\mu$, the chargino amplitude destructively interferes with the SM and
charged Higgs amplitudes, and the full one-loop amplitude can
vanish. However, this cancellation occurs only in a relatively small
region of parameter space. Either case can result in a very large
contribution to $\Delta\chi^2$.  Because of the strong impact of this
observable we have chosen a more conservative error assignment for it
(see Section~\ref{overview}).

As a first example of how the $\mu>0$ region of parameter space is
more severely constrained than the $\mu<0$ region, consider the input
parameters in the supergravity model. In the $\mu>0$ case, the
no-scale model~\cite{Lahanas87} (i.e. $M_0=0$) is seen to be
excluded\footnote{Interesting bounds on $M_0$ and other supersymmetry
parameters have also been obtained by studying charge and color
breaking minima of the effective potential~\cite{Casas96}. However,
these bounds are relaxed when one considers the possibility that the
tunneling time from a false to the true vacuum can exceed the
lifetime of the universe~\cite{Kusenko96}.}.  Specifically, we find
the independent limits $M_0\geq 9$ GeV and $M_{1/2} \gsim 105$~GeV
(Fig.~\ref{pos1}(a)). In the $\mu<0$ case, by contrast, no limit on
$M_0$ can be set, and the limit, $M_{1/2} \gsim 120$~GeV, is not
improved by our analysis.

We should stress at this point that, for either sign of $\mu$, there
is some parameter space excluded even at the very largest values of
$M_0$ and $M_{1/2}$ that we consider (1 and 0.5 TeV,
respectively). Here we emphasize the bounds that are valid
independently of the values of any other parameters. This is what we
mean by ``excluded regions''. For emphasis we sometimes refer to these
regions as ``absolutely excluded''.

The input parameter which sets the scale of the superpartner spectrum
in the gauge-mediated models, $\Lambda$, is constrained in the $\mu>0$
case to be larger than 36 (14) TeV in the GM$_5$ (GM$_{10}$) model,
whereas the entire (quaint) parameter space extends down to 28 (10)
TeV (Figs.~\ref{pos1} (b--c)). Again, the corresponding bound with
$\mu<0$ of 41 (13) TeV could not be improved significantly.

Similarly, Figs.~\ref{pos2} (d--f) show that for $\mu > 0$, $\tan\beta
< 36$, 45 and 39, respectively, for the three models.  In contrast,
for $\mu < 0$ values of $\tan\beta$ as large as 65 are still allowed.

Figs.~\ref{pos1} (d--f) illustrate how the oblique parameters can
serve as useful barometers in parametrizing the excluded region of
supersymmetric parameter space. Here we see, for $\mu>0$, the allowed
and excluded regions in the $S$--$T$ plane.  The region with $T \geq
0.12$ (0.07) is excluded in the SUGRA (GM) model. Also, we can read
off positive and negative bounds on $S$.

The sensitivity of some of the observables is shown in
Figs.~\ref{neg1} (a--c) and \ref{pos1} (g--l). Here we see, for
example, $\delta\sin^2\theta_{\rm eff}^e \leq -6\times10^{-4}$
$(-4\times10^{-4})$ and $\delta A^{FB}(b) \geq 4\times10^{-3}$
$(3\times10^{-3})$ are excluded in the SUGRA (GM) model with $\mu>0$.

The global precision analysis leads to interesting bounds on the
superpartner and Higgs boson masses. In Fig.~\ref{pos2} (a--c) we show
the significantly improved limits, $m_A > 320$, 255, 230~GeV, in the
three models, with $\mu>0$. In the $\mu<0$ case these bounds are
weakened to $m_A > 115$, 220, and 185~GeV (Figs.~\ref{neg1}
(d--f)). These limits imply bounds on other Higgs boson masses. For
example, the region with the lightest possible Higgs boson mass, $62.5
< m_h < 71$~GeV, is absolutely excluded, in every model and for any
sgn($\mu$)\footnote{Table~\ref{bounds} reflects a stronger statement
of this bound based on more recent LEP results, as discussed in the
next section.}. The upper bound on $m_h$ depends on the cut-off chosen
for the squark mass scale.  With an ${\cal O}(1)$ TeV cut-off the
upper limit in the GM models, $m_h\lsim115$ GeV, is stronger by about
10 GeV compared to SUGRA due to the relative absence of mixing in the
top squark sector. All the Higgs-boson mass lower limits primarily
result from the $B(B\rightarrow X_s\gamma)$ constraint on the charged
Higgs mass, e.g. $m_{H^+}>330,\ 265,\ 240$ GeV, in the $\mu>0$ case
(Figs.~\ref{pos2} (d--f)).

The lower bounds on the lightest chargino mass, $m_{\tilde\chi_1^+} >
89,\ 100,\ 120$ GeV, are seen in Figs.~\ref{pos2} (g--i) in the
$\mu>0$ case, together with the corresponding bounds on the lightest
neutralino mass. The indirect bounds on the squark (specifically
$\tilde u_L$) and gluino masses can be read off of Figs.~\ref{pos2}
(j--l) and \ref{neg1} (g--i), and in the $\mu>0$ case extend well
beyond the direct limits. For example, the left-handed up-squark mass
is bounded by 285 (355) GeV in the SUGRA (GM) model.

Both the direct and indirect limits on the third generation squark
masses are different from the first two generations. The indirect
bounds are illustrated in Figs.~\ref{neg2} (a--f) and \ref{pos3}
(a--f).  For a large range of third generation squark masses, in some
cases exceeding 1 TeV, it is possible to find excluded parameter
space. However, the regions that are absolutely excluded are not so
significant for these masses. In particular, the light top squark
cannot be excluded in the SUGRA model (Fig.~\ref{pos3}(d)). However,
the heavy (predominantly left-handed) top squark mass is significantly
bounded by $m_{\tilde t_2} \gsim 275$ (355) GeV in the SUGRA (GM)
model, with $\mu>0$.

The constraints on the slepton masses are shown in Figs.~\ref{pos3}
(g--l) and, for the third generation, in Figs.~\ref{neg2} (g--i). The
absolutely excluded bounds for these and other particle masses and
model parameters are listed in Table~\ref{bounds}. They combine our
present knowledge from phenomenological constraints\footnote{We refer
to electroweak symmetry breaking and Yukawa coupling perturbativity
constraints.}, direct searches and precision experiments. Entries
marked by an asterisk are determined by the phenomenological
constraints or the direct search limits, i.e.\ our global precision
analysis does not improve the bound.

The range of best fit values of $m_t$ is confined to the range
168--174 GeV in the parameter space with $\Delta\chi^2<3.84$. Thus,
supersymmetry prefers $m_t$ in the lower part of the allowed Tevatron
range.  The $Z$ lineshape determination of $\alpha_s$ is known to be
very sensitive to new physics, which in general alters the theoretical
prediction for $R_\ell$.  This happens typically through vertex
corrections or other corrections to the weak mixing angle.  Our
analysis reveals, however, that in the allowed regions of the
investigated models, the fitted $\alpha_s$ values are within the small
range $0.1195 \pm 0.0006$.  Therefore, in the context of minimal
gauge- or gravity-mediated supersymmetry breaking (including the
external constraint $0.118 \pm 0.003$),
$$ \alpha_s = 0.1195 \pm 0.0022 \pm 0.0006, $$ where the first error
is experimental and the second theoretical, i.e.\ from varying the
supersymmetry parameters.

\begin{table}[htb]
\begin{center}
\begin{tabular}{|lc||r@{}l|r@{}l||r@{}l|r@{}l||r@{}l|r@{}l|}
\hline
model                   &   & \multicolumn{4}{c||}{SUGRA} 
                            & \multicolumn{4}{c||}{GM$_5$} 
                            & \multicolumn{4}{c|}{GM$_{10}$} \\
\hline    
sgn ($\mu$) &&+$\;\;$&&$-\;\;$&&+$\;\;$&&$-\;\;$&&+$\;\;$&&$-\;\;$ &  \\
\hline\hline
$M_0$                   &   &    9& &    0&*&  ---& &  ---& &  ---& &  ---&  \\
$M_{1/2}$               &   &  105& &  120& &  ---& &  ---& &  ---& &  ---&  \\
$\Lambda$               &   &  ---& &  ---& &   36& &   41& &   14& &   13&* \\
$|\mu|$                 &   &  135&*&  135&*&  200& &  200& &  180& &  165&  \\
$\tan\beta$             &$>$&  1.4&*&  1.2&*&  1.3&*&  1.2&*&  1.3&*&  1.2&* \\
$\tan\beta$             &$<$&   36& &   60&*&   45& &   65&*&   39& &   52&* \\
\hline
$m_h$                   &   &   78& &   78&*&   78& &   78&*&   78& &   78&* \\
$m_H$                   &   &  320& &  115& &  255& &  220& &  230& &  185&  \\
$m_A$                   &   &  320& &  115& &  255& &  220& &  230& &  185&  \\
$m_{H^+}$               &   &  330& &  140& &  265& &  235& &  240& &  205&  \\
\hline
$m_{\tilde\chi_1^0}$    &   &   45& &   46& &   51& &   50&*&   61& &   48&* \\
$m_{\tilde\chi_2^0}$    &   &   89& &   83& &  100& &   91&*&  120& &   85&* \\
$m_{\tilde\chi_3^0}$    &   &  150&*&  150&*&  215& &  210& &  190& &  175&  \\
$m_{\tilde\chi_4^0}$    &   &  195& &  205&*&  230& &  235& &  225& &  215&  \\
$m_{\tilde\chi_1^+}$    &   &   90& &   80& &  100& &   89&*&  120& &   82&* \\
$m_{\tilde\chi_2^+}$    &   &  195& &  205&*&  235& &  240& &  230& &  220&  \\
$m_{\tilde{g}}$         &   &  255& &  285& &  285& &  320&*&  350& &  335&* \\
\hline
$m_{\tilde e_R}$        &   &   74& &   66&*&   73& &   76&*&   70& &   63&* \\
$m_{\tilde e_L}$        &   &  105& &  110& &  140& &  150& &  125& &  120&  \\
$m_{\tilde\tau_1}$      &   &   53& &   48&*&   73& &   55& &   70& &   52&  \\
$m_{\tilde\tau_2}$      &   &  110& &  110& &  140& &  150& &  125& &  120&  \\
\hline
$m_{\tilde u_L}$        &   &  285& &  295& &  355& &  405&*&  355& &  335&* \\
$m_{\tilde u_R}$        &   &  280& &  290& &  330& &  375&*&  340& &  325&* \\
$m_{\tilde d_L}$        &   &  295& &  305& &  360& &  410&*&  360& &  345&* \\
$m_{\tilde d_R}$        &   &  285& &  290& &  325& &  370&*&  335& &  325&* \\
\hline
$m_{\tilde{t}_1}$       &   &   56& &   66& &  275& &  170&*&  290& &  175&* \\
$m_{\tilde{t}_2}$       &   &  275& &  310& &  355& &  420& &  365& &  385&  \\
$m_{\tilde{b}_1}$       &   &  185& &  250& &  320& &  345& &  330& &  310&  \\
$m_{\tilde{b}_2}$       &   &  285& &  295& &  325& &  370&*&  335& &  330&* \\
\hline
$S$                  &$>$&$-0.02$&&$-0.02$& & 0.00& & 0.00&*&$-0.01$&&0.00&  \\
$S$                     &$<$&+0.04& &+0.07&*&+0.01& &+0.03& & 0.00& &+0.03&* \\
$T$                     &$<$& 0.12& & 0.15& & 0.07& & 0.08& & 0.07& & 0.12&* \\
$U$                     &$<$& 0.03& & 0.05& & 0.02& & 0.03& & 0.02& & 0.05&* \\
\hline
\end{tabular}
\end{center}
\caption{Limits on model parameters and superpartner and Higgs-boson
masses. All masses and mass parameters are in GeV (except for
$\Lambda$ in TeV) and are lower limits.  Limits for $S$, $T$, $U$,
and $\tan\beta$ are as indicated ($T$ and $U$ are positive definite).
Entries marked by an asterisk are consequences of constraints from
electroweak symmetry breaking, Yukawa perturbativity, or direct
searches, and are not significantly improved by our analysis.}
\label{bounds}
\end{table}

\section{Conclusions and Outlook}
\label{con}

Precision data continues to play an important role in discriminating
and constraining various extensions of the standard model. For
example, it is well known that the precision data analysis leads to
serious constraints on technicolor models. Supersymmetric models can
always avoid such constraints because the supersymmetric corrections
decouple, so the supersymmetric models look just like the standard
model if the supersymmetric mass scale is large enough ($\sim0.5$ to
1~TeV).  As long as the standard model with a light Higgs boson
provides a good fit to the data (the current fit is excellent),
supersymmetric models can as well.

On the other hand, what if one were to consider a supersymmetric
spectrum which includes light (${\cal O}(M_Z)$) superpartners? Then it
is interesting to investigate whether the global fits to the data
become better or worse as a result of non-negligible supersymmetric
corrections. This has been the purpose of this paper. To this end, we
combined a state of the art standard model calculation with the
one-loop supersymmetric corrections, and performed global fits to the
precision observables in three supersymmetric models. We have shown
that our global analysis leads (in many cases) to significantly
improved bounds on the parameters and masses of the various
models. For example, we have shown that for a positive $\mu$-parameter
the CP--odd Higgs boson mass lower bound increases from $\sim60$
GeV to over 320 (230) GeV in the supergravity (gauge-mediated)
model. For $\mu < 0$ the $m_A$ bound is not as strong, but we are able
to set a bound,
\begin{equation}
   m_A > 115 \hbox{ GeV},
\label{mabound}
\end{equation}
for all models under consideration. This has an important consequence:
as shown in Figure~21 of the very recent Ref.~\cite{Sopczak97}, there
is unexcluded parameter space with $60 < m_h < 78$ GeV when $68 <
m_A < 102$ GeV. With our bound~(\ref{mabound}) this part of
parameter space is excluded and we conclude,
\begin{equation}
   m_h \geq 78 \hbox{ GeV},
\label{mhbound}
\end{equation}
i.e.\ the MSSM Higgs bound coincides with the SM one. The $m_h$
bound~(\ref{mhbound}) supersedes the bound of 71 GeV mentioned in
Section~\ref{results}, which has a similar origin, but was based on
the data presented at the summer conferences~\cite{AlephHEP97}; for
similar bounds with fixed values of $\tan\beta$ see
Ref.~\cite{Chankowski97}.

We set many other interesting bounds, and our main results can be
found in Table~\ref{bounds} and Figures~\ref{neg1}--\ref{pos3}.  We
advocate Table~\ref{bounds} as a reference for the current collider
and precision bounds in an important class of supersymmetry breaking
scenarios.

In summary, significant portions of the parameter spaces of popular
supersymmetry breaking scenarios have been shown to be excluded, in
particular, for $\mu > 0$. This has immediate consequences for the
near future of collider experiments.  LEP 2 will ultimately be able to
discover a chargino with a mass up to about 100 GeV. Hence, LEP 2 is
very unlikely to discover the chargino in the gauge mediation models
with $\mu>0$. The main injector at Fermilab, a possible further
luminosity upgrade (TEV 33), and the LHC, however, will be able to
cover a large part of parameter space which is not excluded by either
present direct limits or precision experiments.

\section*{Acknowledgements}
We would like to thank Paul Langacker for many valuable comments and
suggestions. We thank Howard Haber, Peter Rowson, and Jim Wells for
useful conversations. We greatly appreciate the hospitality and
stimulating atmosphere of the Aspen Center for Physics and the CERN
theory group. We are thankful for the hospitality of the University of
Washington nuclear and particle theory groups, and thank the computing
center staff at CERN and the University of Washington for support.  We
greatly benefitted from the one-loop integral package, {\tt FF},
written by G.J. van Oldenborgh~\cite{vanOldenborgh90}.

\newpage

\newpage

\newpage

\newpage

\newpage

\newpage

\newpage

\newpage

\newpage

\end{document}